\documentclass[twocolumn]{aastex631}
\usepackage{amsmath}

\begin{document}

\title{The Lower Limit of Dynamical Black Hole Masses Detectable in Virgo Compact Stellar Systems Using the JWST/NIRSpec IFU}

\author[0000-0002-1584-2281]{Behzad Tahmasebzadeh}
\affiliation{Department of Astronomy, 
University of Michigan, 
1085 S. University Ave., 
Ann Arbor, MI 48109, USA}

\author[0009-0009-5509-4706]{Andrew Lapeer}
\affiliation{Department of Astronomy, 
University of Michigan, 1085 S. University Ave., 
Ann Arbor, MI 48109, USA}

\author[0000-0002-5038-9267]{Eugene Vasiliev}
\affiliation{School of Mathematics \& Physics, University of Surrey, Guildford, GU2 7XH, UK}

\author[0000-0002-6257-2341]{Monica Valluri}
\affiliation{Department of Astronomy, 
University of Michigan, 
1085 S. University Ave., 
Ann Arbor, MI 48109, USA}

\author[0000-0003-3009-4928]{Matthew A.\ Taylor}
\affiliation{University of Calgary,
2500 University Drive NW,
Calgary Alberta T2N 1N4, Canada}

\author[0009-0006-7485-7463]{Solveig Thompson}
\affiliation{University of Calgary,
2500 University Drive NW,
Calgary Alberta T2N 1N4, Canada}

\begin{abstract}

Due to observational challenges, the mass function of black holes (BH) at lower masses is poorly constrained in the local universe. Understanding the occupation fraction of BHs in low-mass galaxies is crucial for constraining the origins of supermassive BH seeds. Compact stellar systems (CSSs), including ultra-compact dwarf galaxies (UCDs) and compact elliptical galaxies (cEs), are potential intermediate-mass BH hosts. Despite the difficulties posed by their limited spheres of influence, stellar dynamical modeling has been effective in estimating central BH masses in CSSs. Some CSSs may harbor a BH constituting up to 20\% of their host stellar mass, while others might not have a central BH.   
In support of our ongoing efforts to determine the BH masses in select CSSs in the Virgo cluster using JWST/NIRSpec IFU observations and orbit-superposition dynamical models, we create mock kinematic data mimicking the characteristics of observed cEs/UCDs in the Virgo cluster with different BH masses. We then construct a series of dynamical models using the orbit-superposition code FORSTAND and explore the accuracy of recovering the BH mass. We find that the mass of BHs comprising 1\% or more of the total host stellar mass can be accurately determined through kinematic maps featuring higher-order velocity moments. We also assess how BH mass measurement is affected by deprojection methods, regularization factors, anisotropy parameters, orbit initial conditions, the absence of higher-order velocity moments, spatial resolution, and the signal-to-noise ratio.



\end{abstract}

\keywords{Stellar Dynamics (1596) --- Supermassive black holes (1663) --- Ultracompact dwarf galaxies (1734) ---
Galaxy evolution(594) ---
Compact galaxies(285) ---}

\section{Introduction} \label{sec:intro}

Compact elliptical (cE) galaxies are identified as a class of compact stellar systems (CSSs) with radii from a few hundred parsecs to about a few thousand parsecs. Despite their diminutive size, they possess significant mass ranging from $10^{8}$ to $10^{10}\;M_{\odot}$.
The stars within these galaxies are predominantly old, indicating that most of their star formation occurred in the distant past \citep{Chilingarian.2009S}.

Ultracompact dwarf galaxies (UCDs) are recognized as another class of CSSs that exhibit properties intermediate between those of typical dwarf galaxies and globular clusters (GCs). They were first identified in studies by \cite{Hilker.1999,Drinkwater.2000}. UCDs are characterized by half-light radii on the order of a few tens of parsecs and masses ranging from approximately $10^{6}$ to $10^{8}\;M_{\odot}$ \citep{Brodie_2011}.

The formation of both cEs and UCDs remains an ongoing debate. cEs found near a larger host galaxy tend to exhibit redder colors, reduced sizes, and older ages compared to isolated ones. These ranges of observed properties reinforce the possibility of various pathways for the formation of cEs. Some may develop gradually by accumulating stellar mass in an isolated environment \citep{Zolotov.2015}, whereas others might originate from the tidal stripping of a galaxy by a more massive nearby galaxy \citep{Bekki.2001, Deeley.2023}.

Two primary formation mechanisms have been proposed for UCDs. The first suggests that UCDs are the remnants of dwarf galaxies that have undergone major disruption, losing most of their stellar mass and leaving behind only their nuclei \citep{2003Ap&SS.285..113G, 2003MNRAS.344..399B, 2008MNRAS.389..102T, 2021MNRAS.506.2459M, mayes_2024}. The second theory proposes that UCDs could be exceptionally massive, outlier members of the GC population possibly formed during intense gas-rich mergers of young massive clusters \citep{Fellhauer.2002,Mieske.2006,mieske_2012}. 

In the stripped nuclei formation channel, cEs/UCDs could retain their primary central supermassive black holes (SMBHs). A key trait of UCDs is their elevated dynamical mass-to-light ratio, indicating the presence of an additional unseen mass component, potentially a central SMBH. It has been shown that roughly two-thirds of UCDs with mass $> 10^7 \; M_{\odot}$ and one-fifth of UCDs with masses in the range of $2\times 10^6 - 10^7\; M_{\odot}$ require an additional mass component \citep{mieske2013, voggel2019}. Considering the presence of an unseen mass component within a portion of the UCD population and the possibility of a stripped nuclei formation scenario for UCDs, it follows SMBHs may reside in a significant fraction of UCDs. 

Stellar dynamical modeling approaches have been successful in constraining the masses of SMBHs within several cEs and UCDs utilizing data from integral field spectroscopic instruments. SMBHs have been detected in nearby cEs, M 32 \citep{Verolme.2002}, NGC 404 \citep{Seth.2010}, and an intermediate mass BH in NGC 205 \citep{Nguyen.2019}. The first confirmation of a SMBH within a UCD was reported by \citet{seth2014}, who measured a central black hole mass of $2.1^{+1.4}_{-0.7}\times 10^{7}\; M_{\odot}$ in M60-UCD1. More recently, 
SMBHs ave been detected in three more UCDs in the Virgo cluster \citep{ahn2017,Ahn.2018} and in a single UCD in the Fornax cluster \citep{Afanasiev.2018}. These findings have laid the groundwork for exploring the formation of cEs/UCDs, and understanding the demographics of SMBHs in the local universe.

Hydrodynamical simulations, particularly the EAGLE project \citep{crain_eagle,schaye_eagle}, support the presence of SMBHs in a significant fraction of UCDs. \citet{mayes_2024} found that about $\sim 51\%$ of stripped nuclei UCDs with masses over $M_{\mathrm{tot}} > 2\times10^6 \; M_{\odot}$ contain SMBHs above $M_{\mathrm{BH}} > 3\times10^5 \; M_{\odot}$. These findings align with the elevated mass-to-light ratio observed in UCDs \citep{mieske2013}. Such results imply that a substantial fraction of the SMBHs population in galaxy clusters remains undetected.

With the introduction of over 600 previously undetected Virgo UCD candidates by the Next Generation Virgo Cluster Survey (NGVS) \citep{liu_ngvs}, and the capabilities of the JWST/NIRSpec IFU, the study of the SMBH population within cEs/UCDs is more feasible than ever.

We aim to apply stellar dynamical modeling to determine the masses of central BHs in a subset of CSSs in the Virgo cluster, utilizing data from JWST/NIRSpec IFU.  In this paper, we theoretically evaluate the capabilities and limitations of the stellar dynamical modeling approach in constraining the central BH masses in Virgo cluster cEs/UCDs.
 

We employ the \citet{sch_1979} orbit-super\-position method as implemented in the FORSTAND code \citep{Vasiliev.2020A}, which is included in the AGAMA stellar-dynamics toolbox \citep{agama}. This code has been already used for measuring the central BH masses in two disk galaxies \citep{Roberts.2021,Merrell.2023}. 

We construct  $N$-body realizations of cEs/UCDs-like systems and use them to generate mock kinematic datasets with properties similar to those in the Virgo cluster and detectable with JWST/NIRSpec. We then utilize orbit-superposition modeling to establish a potential lower limit for the detectable central BH mass in Virgo CCSs. Section  \ref{sec:2} provides a detailed overview of how we construct the mock sample and generate mock kinematics. In section \ref{sec:3}, we delve into the Schwarzschild modeling applied to mock cEs/UCDs with varying BH mass fractions. In section \ref{sec:4}, we present our results and discuss the limitations of our modeling approaches for such CCSs.

\section{Mock Data} \label{sec:2}

\begin{figure*}
	\centering	%
	\includegraphics[width=2\columnwidth]{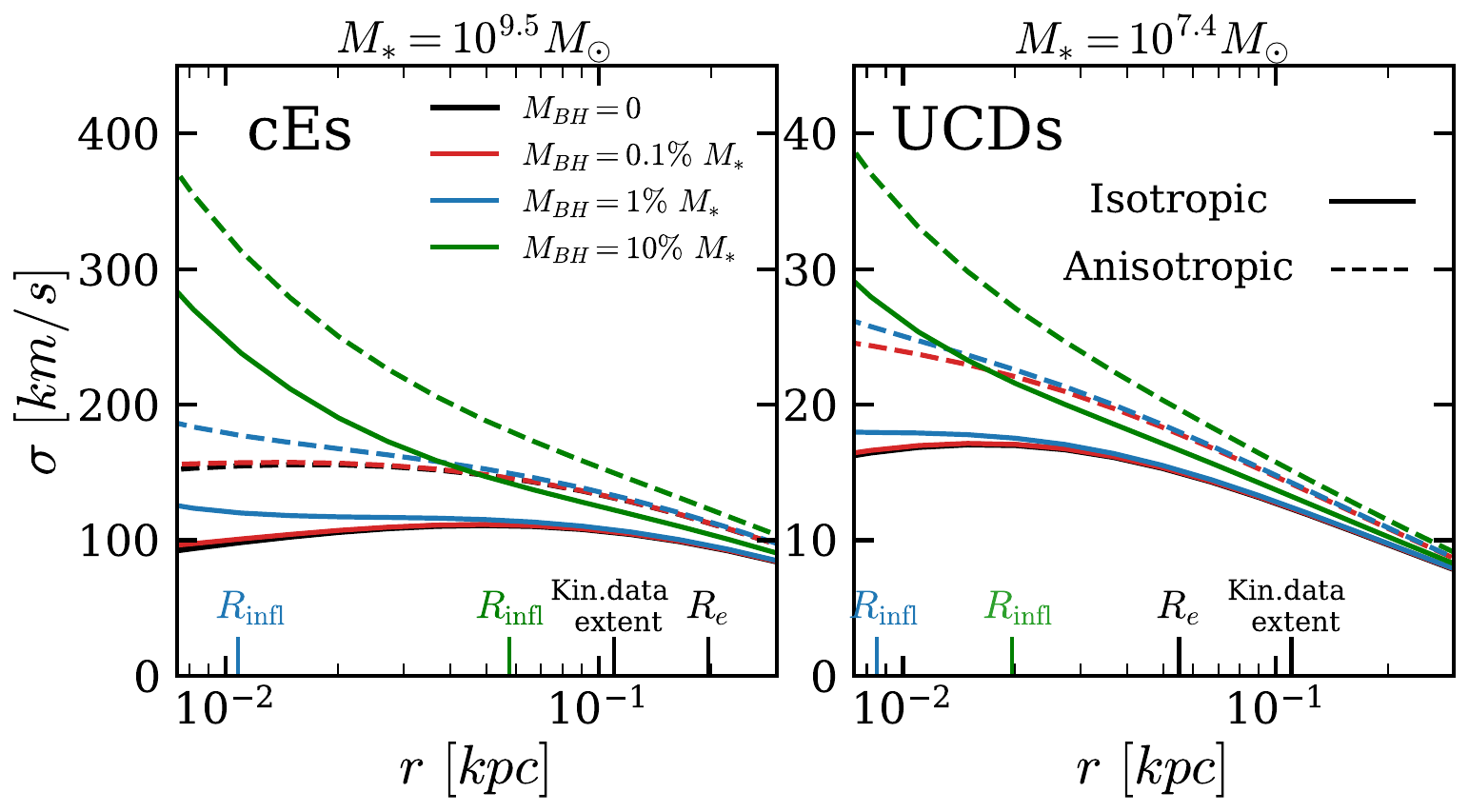}
	\hspace{8pt}%
	\caption{Intrinsic velocity dispersion profiles of mock cEs (left) and mock UCDs (right) are presented for different black hole masses: $M_{BH} = 0$ (black), $M_{BH} = 0.1\% M_{*}$ (red),  $M_{BH} = 1\% M_{*}$ (blue), and $M_{BH} = 10\% M_{*}$ (green). Solid and dashed lines indicate isotropic $(\beta = 0)$, and anisotropic models $(\beta = 0.5)$ models, respectively. On the horizontal axis, markers indicate the mock kinematic spatial scale, the sphere of influence for each model, the half-light radius, and the extent of the mock kinematic data.}%
	\label{fig:Sigma}%
\end{figure*}

Our mock data sets  are generated from  models  with $10^7 < (M_*/M_\odot) < 10^{9.5}$ and $50<(r_{e}/\mbox{pc})<300$, where $M_*$ is the stellar mass and $r_{e}$ is the half-light radius. 

\subsection{Generating N-body snapshots} \label{mock_gen}
We employed the AGAMA stellar dynamics framework to create self-consistent N-body distributions representing UCDs/cEs. Each model consists of a spherical stellar component and a central black hole. 
The majority of UCDs that have been confirmed through spectroscopic analysis exhibit a nearly spherical shape \cite[]{Zhang.2015}. Therefore, we use a spherical double-power-law stellar density profile of the form: 
\begin{equation} \label{eq:1}
\rho(r) = \rho_0 \Big( \frac{r}{a} \Big)^{-\gamma}
\bigg[1+\Big( \frac{r}{a} \Big)^{\alpha}\bigg]^{\frac{\gamma-\beta}{\alpha}} 
\exp\bigg[-\Big(\frac{r}{r_{\mathrm{cut}}}\Big)^{2}\bigg],
\end{equation}
where $\gamma$ is the inner power-law slope, $\beta$ is the outer power-law slope, $\alpha$ is the steepness of transition between these asymptotic regimes, $a$ and $r_{\mathrm{cut}}$ are the scale and cutoff radii of the system, and $\rho_0$ is the density normalization.
The black hole is represented by a Plummer potential with a mass $M_{\rm BH}$ and a very small softening radius $10^{-4}$ kpc. 

For the given density and potential profiles, we determine the spherical anisotropic distribution function (DF) using the \citet{Cuddeford.1991} method. In terms of energy $E$ and angular momentum $L$, this DF takes the following form:
\begin{equation} \label{eq:3}
    f(E,L) = \hat{f}(E)\; L^{-2\beta_0},
\end{equation}
where, $\beta_0 \equiv 1 - \frac{\sigma_t^2}{\sigma_r^2}$ is the velocity anisotropy coefficient, in which $\sigma_t$ and $\sigma_r$ are the tangential and radial velocity dispersions, and $\hat{f}(E)$ is computed numerically. 

We create an $N$-body representation of each mock model by sampling phase-space coordinates $\boldsymbol x, \boldsymbol v$ for $N=10^6$ equal-mass particles from the DF.

Previous studies did not find evidence of dark matter (DM) in UCDs \cite[]{Frank.2011, Strader.2013}.  During the tidal stripping of nucleated dwarf galaxies, it is expected that most of the outer DM halo will be stripped away first \citep{Smith.2016}. 
The presence of DM in cEs is still debated. While some studies have found a significant amount of DM in cEs, others have not found evidence for DM, suggesting that different cEs may have different formation pathways \cite[]{Buote.2019, Yildirim.2017}.

For simplicity, in the main body of this study, we do not include a DM halo in our mock models. However, in Appendix \ref{appen}, we examine a cE model that includes a DM halo. We demonstrate that cored or cuspy  Navarro–Frenk–White (NFW) DM halo \cite[]{NFW.1996} does not effect the central kinematics of cEs, a critical factor in precisely determining BH masses.

\subsection{Mock Sample}

We consider two stellar mass distributions both described by the density profile in equation~\ref{eq:1}. The model for the UCD has  $a=50$ pc, $\gamma=1$, $M_* = 10^{7.4}$ $ M_{\odot}$,  $\beta = 4$, $\alpha = 1$, $r_{\mathrm{cut}} = 12a$, which corresponds to $r_{e}\sim 63$ pc, and  $n=2.23$. The model for the cE has $a=330$ pc, $\gamma=1.5$, $M_* = 10^{9.5}$ $ M_{\odot}$,  $\beta = 4$, $\alpha = 1$, $r_{\mathrm{cut}} = 3a$, 
corresponding to $r_{e}\sim 245$ pc and $n=3.07$. 

For each UCD and cE mock model, we adopt two DFs given by equation~\ref{eq:3}: one isotropic ($\beta_0=0$) and the other a moderately radially anisotropic ($\beta_0=0.5$) . 
We include a BH in each model with three possible masses of $M_{\rm BH}/M_* = 10$\%, 1\%, and 0.1\%. Thus, we have 12 mock models in total, which will be used to examine our dynamical modeling approach.  

Fig.~\ref{fig:Sigma} represents the intrinsic velocity dispersion profiles of our 12 mock models (solid lines are isotropic models, and dashed lines indicate anisotropic ones) compared to a similar isotropic mock model but without a BH (black line). Markers on the horizontal axis indicate: the spatial scale of the mock IFU kinematic data; the BH sphere of influence ($R_{\mathrm{infl}}$), defined as the radius at which the enclosed mass of stars equals the mass of the BH; the half-light radius ($r_{e}$); and the extent of the mock IFU kinematic data. 

The figure shows that in both cEs and UCDs, BH masses that constitute $1\%$ or less of the stellar mass ($M_*$) do not significantly affect the intrinsic velocity dispersion within the central regions. 

The impact on velocity dispersion is much more noticeable when the BH mass ($M_{\rm BH}$) is at $10\%$ of $M_*$. This effect is more pronounced in anisotropic models as compared to isotropic ones. For BH masses below $1\% M_*$, the $R_{\rm infl}$ is smaller than the spatial resolution achievable in JWST/NIRSpec-based kinematic datasets at the distance of Virgo for both cEs and UCDs.
We found that at certain radii, the velocity dispersion in an isotropic cE/UCD with $M_{BH}=10\% M_*$ can match that of an anisotropic cE/UCD with $M_{BH}=1\% M_*$. However, the velocity dispersion profile gradients in the inner regions differ significantly, with a larger BH mass forecasting steeper velocity dispersion profiles.


\subsection{Generation of mock photometry and kinematics} \label{subsec:3A}
To create mock images, we place the models at the distance of the Virgo cluster at $16$\,Mpc. At this distance $1\arcsec \simeq 82\,\mathrm{pc}$. The mock images are created with a spatial scale of $ 0.05 $ arcsec pixel$^{-1}$ similar to HST (ACS) images and extended to 6 arcsec for UCDs and 25 arcsec for cEs. We then convolved the mock images with a PSF defined by one component circular Gaussian with FWHM of $\sim 0.1$ arcsec. HST images typically have PSF sizes ranging from approximately $\sigma \sim$ 0.07 to 0.1 arcsec, depending on the selected filter and camera channel.

To construct kinematic data sets meant to simulate those from
JWST/NIRSpec IFU, we 
create mock kinematic data extended to $1.5$ arcsec  with $0.1$ arcsec spatial resolution (note that $r_{e}^{\rm UCDs}\sim 0.75$ arcsec, and $r_{e}^{\rm cE}\sim 2.9$ arcsec).
We first bin particles lying within a projected radius of $1.5$ arcsec from the center of the galaxy into pixels of $0.1\arcsec \times 0.1\arcsec$. We then apply Voronoi binning \cite[]{Cappellari.2003} with the target signal-to-noise ratio threshold of $S/N = 35$ (Using the number of particles per bin divided by the Poisson error as a proxy) which leads to $N_{\rm kin} \sim 60$ apertures for each kinematic map. For JWST/NIRSpec IFU, the PSF FWHM can vary across the near-infrared spectrum from about 0.1 to 0.2 arcsecond or more, depending on the exact wavelength and observing conditions. We thus convolved the apertures with a Gaussian PSF with FWHM of $\sim 0.11$ arcsec. Then, we convert the line-of-sight velocity distributions (LOSVD) into the Gauss-Hermite (GH) representation with 4 moments $(v_{o}, \sigma_{o}, h_{3}, h_{4})$ \cite[]{Marel.1993, Gerhard.1993}.

To create mock kinematic errors, we use a logarithmic function that correlates with the number of pixels per Voronoi bin inferred from \citet[]{Tsatsi.2015}, then normalized it to be in the range of mean value of 2.5 km/s for $v_{o}$ and $\sigma_{o}$ maps, and 0.02 for $h_{3}$ and $h_{4}$ maps. To have realistic noisy data, we perturb the kinematic data by adding Gaussian random noise with amplitude specified by the error maps.  

\begin{figure*}
	\centering	%
	\includegraphics[width=\columnwidth]{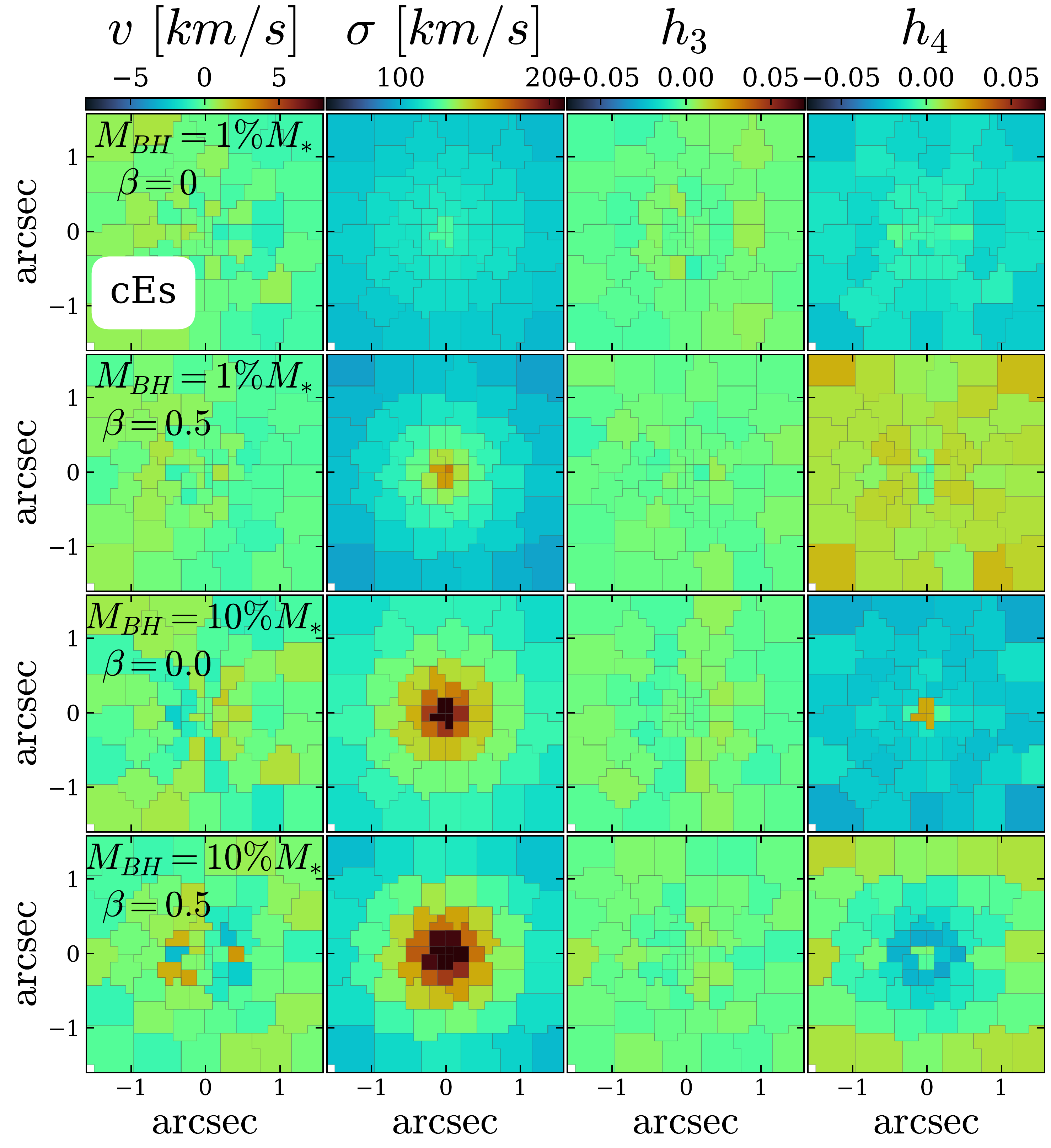}
 	\includegraphics[width=\columnwidth]{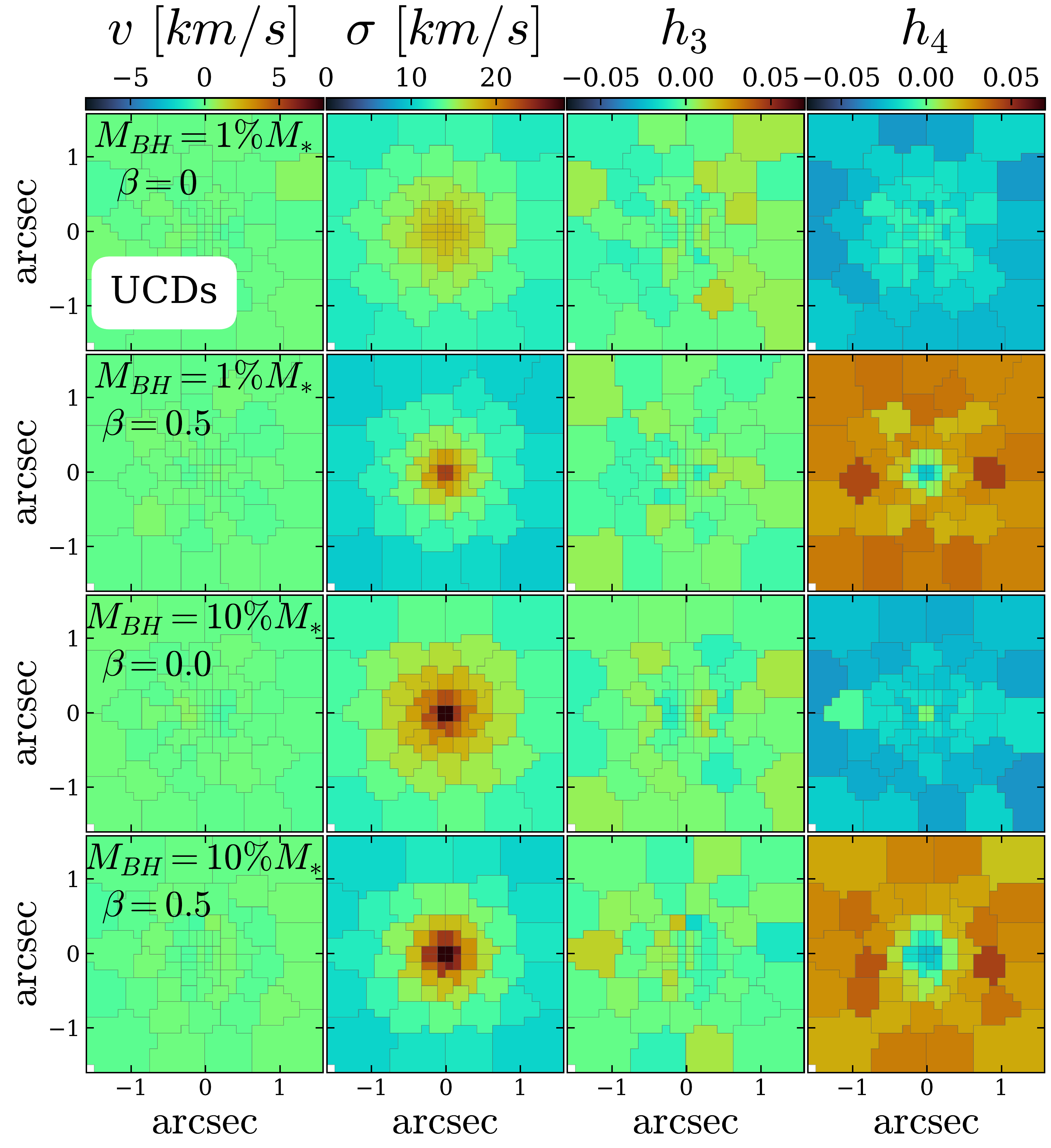}
  \hspace{8pt}%
	\caption{The noise-free kinematic maps for a set of mock cEs (left panel) and a set of mock UCDs (right panel) for different BH masses and anisotropy coefficients. The columns from left to right are the maps of GH coefficients ($v_{o}$, $\sigma_{o}$, $h_3$,  $h_4$).}%
	\label{fig:kinem}%
\end{figure*}

Fig. \ref{fig:kinem} shows the noise-free kinematic maps for mock cEs (left panel) and mock UCDs (right panel). The rows from top to bottom indicate mock models with different BH masses and anisotropy coefficients as ($M_{\rm BH} = 1\% M_*$,  $\beta_0=0$), ($M_{\rm BH} = 1\% M_*$,  $\beta_0=0.5$), ($M_{\rm BH} = 10\% M_*$,  $\beta_0=0$), and ($M_{\rm BH} = 10\% M_*$,  $\beta_0=0.5$), respectively. 

The anisotropic models demonstrate a significantly higher $\sigma_{o}$ in the inner regions, which decreases more steeply outward when compared to isotropic models. Furthermore, anisotropic models exhibit a notably steeper outward increase in $h_{4}$ values, resulting in $h_{4}$ tending to be positive in their outer regions. In isotropic models, the LOSVD in the central bins is slightly sharper and narrower compared to anisotropic mock data. But in the outermost bins, although the widths of distribution in both models are similar, the anisotropic mock data exhibits a LOSVD with a significantly more pronounced peak.

\section{Schwarzschild Dynamical Modeling} \label{sec:3}

\subsection{Inferring the 3D luminosity density}
We employ two different methods to infer the 3D luminosity density of the mock images. The density forms we use for deprojection differ from that employed in constructing the N-body models used to generate the mock kinematic maps (Eq. \ref{eq:1}). The different density forms are deliberately adopted to examine the robustness of our dynamical modeling approach.

First, we apply the multi-Gaussian expansion (MGE)\citep[]{cappellari.2002} to parametrize the surface brightness of the mock image, then we deproject the 2D MGEs to 3D spherical MGEs.  


In the second method, we use GALFIT \cite[]{Peng.2010} to fit a S\'{e}rsic function to describe the surface brightness of our mock image. Then, the 3D density is obtained by numerically computing the deprojection integral (which is well-defined in any spherical system) on a grid in radius and constructing an interpolating spline. The 3D mass density profile is obtained by multiplying a constant stellar mass-to-light ratio $M/L$ with the 3D luminosity density. 


\subsection{Gravitational Potential}
We construct the total gravitational potential, including contributions from both the stars and a central BH. The potential of the stellar component is derived from the inferred 3D stellar luminosity density distribution (discussed in the previous subsection), using an assumed $M/L$ and by solving the Poisson equation.  The potential of the black hole is represented as a Plummer potential with a fixed scale radius of $a=10^{-4}$ kpc. Thus, our models have two free parameters: $M/L$ and $M_{\rm BH}$.

\subsection{Construction of orbit library} \label{IC_detail}

Unlike other Schwarzschild modeling codes, which assign orbit initial conditions for the orbit library on a regular grid in the space of integrals of motion, the FORSTAND code samples the orbital initial conditions randomly in the 6d phase space (see Section~2.4 in \citealt{Vasiliev.2020A} for a discussion). Namely, the positions are sampled uniformly from the intrinsic stellar density profile of the model. Velocities are then drawn from a Gaussian distribution with position-dependent dispersions obtained by solving the anisotropic Jeans equation for the axisymmetrized potential and density (e.g., \citealt{Cappellari.2008}. 
We usually set the anisotropy parameter $\beta_0$ to zero; however, in the next section we also examine the impact of initial conditions generated with a non-zero anisotropy parameter on the modeling outcomes.

For every set of model parameters ($M_{\rm BH}$ and $M/L$), we build an orbit library by integrating $N_{\rm orb}=$ 20\,000 orbits over 100 dynamical periods within the specified potential. The LOSVD of each orbit is first recorded as three-dimensional data cubes, which consist of two coordinates in the image plane and the velocity axis. These are represented in terms of a basis set of tensor-product B-splines with a degree of 2. Following this, they are convolved with the spatial PSF and re-binned onto the array of Voronoi apertures. 

\begin{figure*}
	\centering	%
	\includegraphics[width=0.65\columnwidth]{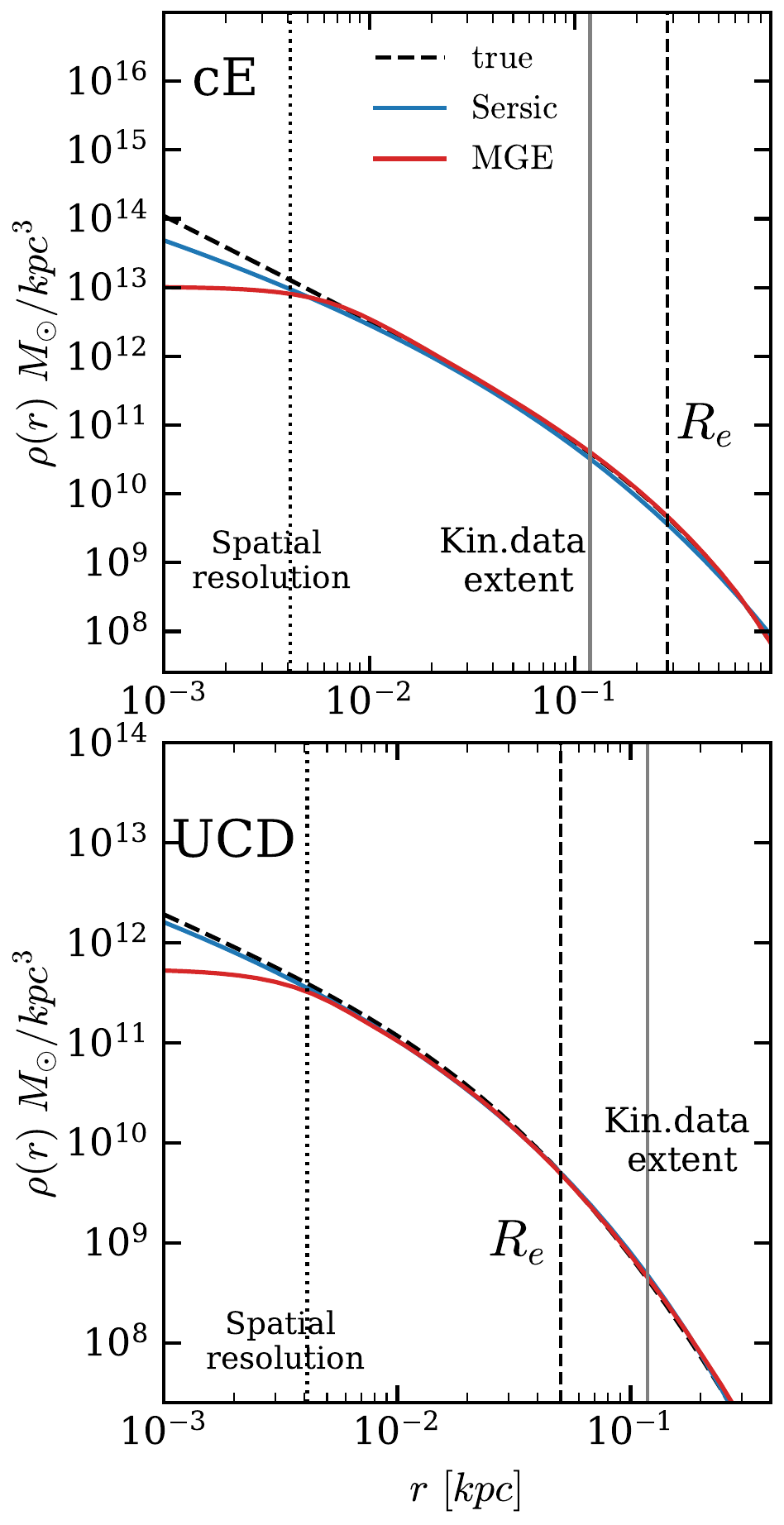}
 	\includegraphics[width=1.3\columnwidth]{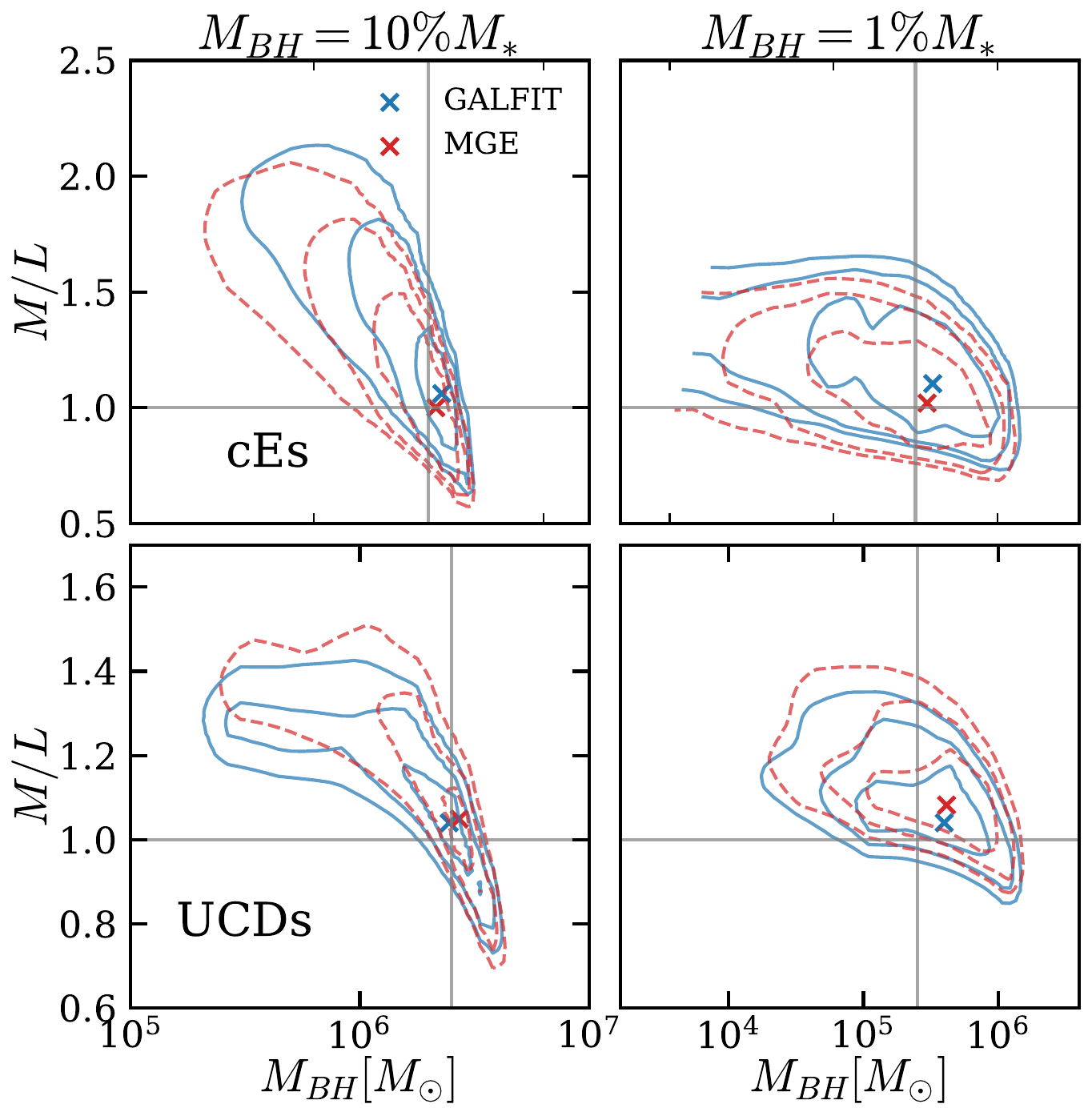}
	\hspace{8pt}%
	\caption{First column: density profile of mock cEs (top) and mock UCDs (bottom), plotted for true density (black dashed line), deprojected model using Sersic parameterization (blue), deprojected model using MGE parameterization (red). Second and third columns: $\chi^2$-contours for best-fit models on the parameter grid of black hole mass versus stellar mass-to-light ratio $(M/L)$ for cEs (top) and UCDs (bottom) with $M_{BH} = 10 \% M_{*}$, $M_{BH} = 1\% M_{*}$, respectively. The contours indicate the $1\sigma$, $2\sigma$, and $3\sigma$ uncertainties ($\Delta \chi^{2} =  2.3, 6.2, 11.8$) for the best-fitting models using different deprojected density models. The crosses of each color indicate the best-fitting model for each data set, and vertical and horizontal gray lines indicate the true values of $M/L$ and $M_{BH}$, respectively.}%
	\label{fig:deproj}%
\end{figure*}

\subsection{Parameter grids and fitting procedure} 
After constructing the orbit library, we find the orbital weights that (a) reproduce the 3D density discretized over a cylindrical grid of $20 \times 15$ in the $R$, $z$ plane, and (b) minimize the objective function $\mathcal{F} \equiv \mathcal{F}_{\text {kin }}+\mathcal{F}_{\text {reg}}$. The first term $\mathcal{F}_{\text {kin }}$ determines the goodness of fit to the kinematic constraints $(v_{o}$, $\sigma_{o}$, $h_3$, $h_4)$. The second term $\mathcal{F}_{\text {reg}}$ is the `regularization-term', which penalizes large differences between orbital weights $\boldsymbol{w}$ to avoid overfitting. We define $\mathcal{F}_{\text {reg }}=\lambda N_{\text {orb }}^{-1} \sum_{i=1}^{N_{\text {orb }}}\left(w_i / \bar{w}\right)^2$, where the mean orbit weight is $\bar{w} \equiv M_{\star} / N_{\text {orb }}$, and the regularization coefficient $\lambda$ controls the trade-off between accurately reproducing kinematic constraints (with a smaller $\lambda$) and the smoothness of the model (with a larger $\lambda$).

Upon identifying the optimal orbit weights, we calculate the final $\chi^{2}$, which assesses the fit relative to the originally measured values $(v_{o}, \sigma_{o}, h_3, h_4)$ and their corresponding uncertainties. This calculation differs slightly from $\mathcal{F}_{\text{kin}}$. However, it is noteworthy that both the $\chi^{2}$ and $\mathcal{F}_{\text{kin}}$ functions exhibit similar shapes and minima locations when plotted against the model parameters.

We construct different orbit libraries on a grid of models that spans a wide range of $M_{\rm BH}$ from 0 to $M_{\rm BH}=90 \% M_{*}$. Consequently, each orbit library is utilized multiple times to explore a range of $M/L$ values by multiplying the velocities by $\sqrt{M/L}$ in multiplicative steps of 0.01 until the minimum of $\chi^{2}$ is found and bracketed from both ends.

As our first attempt, we set $\lambda=1$, which is adequate (in our case) to prevent overfitting and to ensure a reasonably smooth likelihood surface when using full kinematic information. However, we will explore the effects of different $\lambda$ values on the modeling outcomes in the next section. 

\section{Results} \label{sec:4}

\subsection{ Models with MGE and S\'{e}rsic parametrization} 

 \begin{figure*}
	\centering	%
	\includegraphics[width=2\columnwidth]{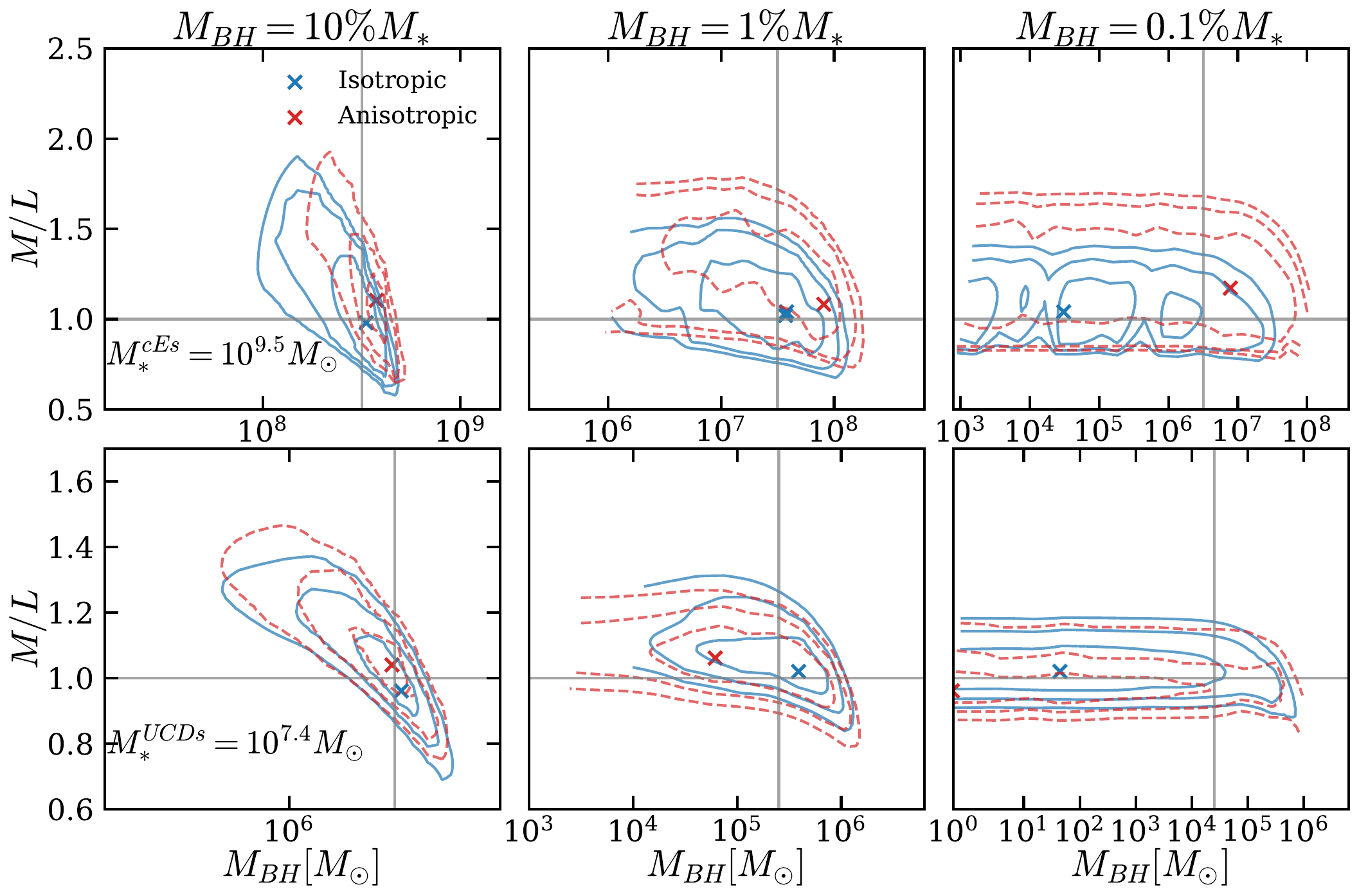}
	\hspace{8pt}%
	\caption{$\chi^2$ contours for the best-fit models on the parameter grid of black hole mass versus the stellar mass-to-light ratio $(M/L)$ plotted for cEs (top) and UCDs (bottom) with $M_{BH} = 10 \% M_{*}$, $M_{BH} = 1\% M_{*}$, and $M_{BH} = 0.1\% M_{*}$, respectively. The contours indicate the $1\sigma$, $2\sigma$, and $3\sigma$ uncertainties for the best-fitting models using isotropic (blue) and anisotropic mock models (red). The cross indicates the best-fitting model for each data set, and vertical and horizontal gray lines indicate the true values of $M/L$ and $M_{BH}$, respectively.}%
	\label{fig:chi2}%
\end{figure*}

As discussed in \citet{Roberts.2021}, various methods of surface brightness parametrization can yield significantly different profiles within the very inner region (e.g., less than image spatial resolution $r<10^{-1}$ arcsec), which is crucial for measuring black hole (BH) mass. 

The innermost part of the S\'{e}rsic profile tends to be more cuspy. The use of such a profile can result in a higher stellar mass in the very central regions and consequently lead to a lower recovered BH mass. In contrast, the MGE parametrization can result in a flat-core profile within the innermost region depending on the PSF size.


However, the variation in dynamical modeling results due to a central cuspy or cored profile is dependent on the spatial resolution of both the photometric image and IFU data cube. We expect that if the IFU's spatial resolution is comparable to or exceeds the photometric spatial resolution, any discrepancies in density profiles at scales within the photometric resolution will not significantly impact the accuracy of BH mass estimations. 

The first column in Fig. \ref{fig:deproj} shows the 3D density profile of mock cEs (top) and mock UCDs (bottom). Plotted are the true (input) density profile (black dashed curve), the deprojected density model using S\'ersic parameterization (blue), and the deprojected density model using MGE parameterization (red). The MGE parameterization results in a cored profile in regions smaller than the spatial resolution ($r< 0.06$ arcsec). The S\'{e}rsic profile more closely approximates the true density within the spatial resolution (shown by a vertical dotted line). In the following, we will explore whether and how these differences can influence the BH mass estimate obtained.

The spatial resolution of our mock kinematic data is $ 0.1 $ arcsec, similar to that expected from the JWST/NIRSpec IFU.  This is also comparable to the spatial resolution (PSF FWHM) of our mock images and $\gtrsim 2\times$ larger than spatial scale (pixel size) 
of the mock images. 

The second and third columns in Fig. \ref{fig:deproj} illustrate the model parameter grid for black hole mass versus the stellar mass-to-light ratio $(M/L)$ plotted for the isotropic cE (top) and the UCD (bottom) with $M_{BH} = 10 \% M_{*}$, $M_{BH} = 0.1\% M_{*}$, respectively.  The contours indicate the $1\sigma$, $2\sigma$, and $3\sigma$ uncertainties ($\Delta \chi^{2} =  2.3, 6.2, 11.8$) for the best-fitting models  
obtained by deprojected density using Sersic (blue), and MGE parameterization (red). The cross marks the best-fitting model for each dataset, while the vertical and horizontal gray lines denote the true values of $M/L$ and $M_{BH}$, respectively. 

For these mock cEs and UCDs, modeling with deprojected density inferred from Sersic and MGE parameterizations does not yield a significant difference in the recovered BH mass. Despite some degeneracy between $M/L$ and $M_{BH}$, their best-fit values are close to the true values. In both models with $M_{BH} = 10\% M_{*}$, the $M/L$ and $M_{BH}$ are very well recovered. In cases where $M_{BH} = 1\% M_{*}$, although the best-fit values lie within the 1$\sigma$ contours, the recovered $M_{BH}$ is slightly higher than the true values.  As we will show in the next section, when modeling mock UCDs/cEs with $M_{BH} = 0.1\% M_{*}$, it is not possible to accurately constrain the BH mass; we can only establish upper limits for the BH mass.

We conclude that in our mock UCDs/cEs data sets, BHs with masses at least a few percent ($> 1\%$) of the host stellar mass can be well recovered regardless of whether the stellar distribution is modeled by a S\'ersic profile or by MGE.

\subsection{Isotropic versus anisotropic models}
In addition to examining the impact of varying BH masses, we also explore how different intrinsic velocity profiles affect the recovery of BH mass through isotropic and anisotropic mock models. Fig. \ref{fig:chi2} represents the model parameter grid for the stellar $M/L$ versus the BH mass plotted for mock cEs (top) and UCDs (bottom), with $M_{BH} = 10 \% M_{*}$, $M_{BH} = 1\% M_{*}$, and $M_{BH} = 0.1\% M_{*}$. The contours indicate the $1\sigma$, $2\sigma$, and $3\sigma$ uncertainties for the best-fitting models using isotropic (blue) and anisotropic mock models (red). 

In all scenarios, the true values are within (1--2)$\sigma$ contours, however, they are tightly constrained only in the models with $M_\mathrm{BH}=$10\%$M_*$.  In the case of an intermediate-mass BH with $M_\mathrm{BH}=$1\%$M_*$ the BH mass is marginally well recovered in the isotropic models, but less accurate in the case of the anisotropic models. When $M_\mathrm{BH}=$0.1\%$M_*$, both isotropic and anisotropic models only allow us to establish upper limits for the BH mass.

\subsection{Modeling without $h_{3}$ and $h_{4}$} 
Dynamical modeling of galaxies may suffer from the mass--anisotropy degeneracy problem \citep{Binney.1982, Merritt.1987}. 
The degeneracy stems from the fact that the observed velocity dispersion alone may not be sufficient to distinguish between two distinct scenarios: one in which a galaxy contains stars moving in radial orbits within a relatively shallow gravitational potential, and another where stars follow more circular orbits within a deeper gravitational potential. To overcome this limitation, it is essential to incorporate higher-order moments of the LOSVD $h_{3}$ and $h_{4}$ as constraints in the modeling \citep[e.g.,][]{Merrifield&Kent.1990, Marel.1993, Gerhard.1993}.

For some of our targets and as a result of observational challenges associated with the spectral resolution of JWST/NIRSpec IFU, it may not always be possible to extract higher-order moments of the LOSVD. Moreover, the majority of dynamical modeling studies rely on standard Jeans equations, which do not use any higher-order moments. 

\begin{figure*}
	\centering	
	\includegraphics[width=2\columnwidth]{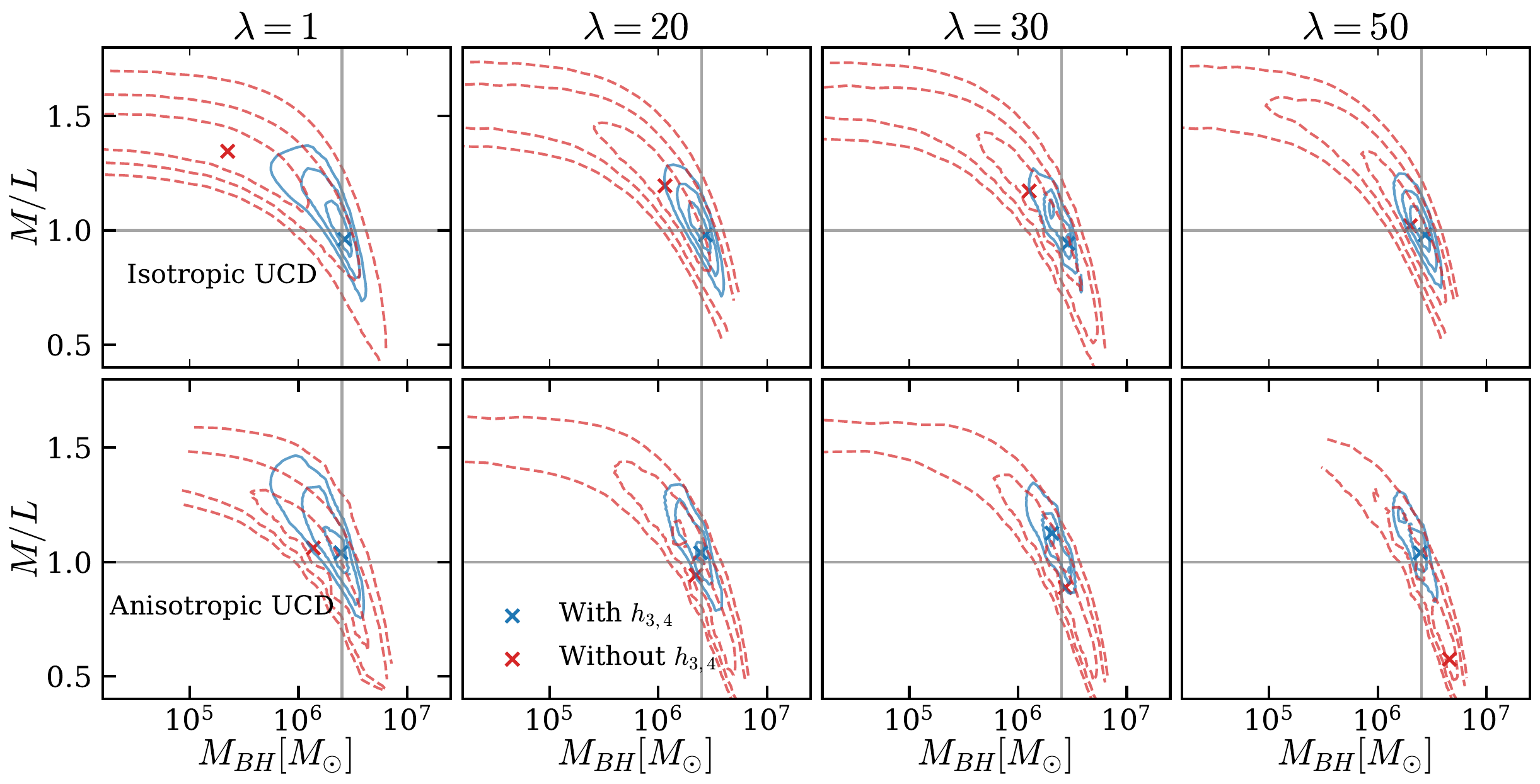}
	\caption{Comparing the results of modeling the
 isotropic UCD (top row) and anisotropic UCD (bottom row) with $M_{BH} = 10 \% M_{*}$, using higher-order GH moments $h_3$, $h_4$ (blue contours) and using only $v$, $\sigma$ (red contours). The columns from left to right show the results of models with different regularization factors of $\lambda=1, 20, 30$, and $50$ respectively.}%
	\label{fig:chinoh34}%
\end{figure*}

Therefore, we reran all models without $h_{3}$ and $h_{4}$ constraints to address this issue under different scenarios and explore potential solutions. We found that the recovered BH mass is significantly biased in most cases, allowing only for an upper-limit estimation. The first column in Fig. \ref{fig:chinoh34} compares the result of the modeling of isotropic UCD (top panel) and anisotropic UCD (bottom panel) with a BH of $M_\mathrm{BH}=$10\%$M_*$. The models using $h_{3}$ and $h_{4}$ constraints are shown in blue contours, while those without these constraints are depicted in red contours.

The bias is even more significant for models with $M_\mathrm{BH}=$1\%$M_*$ and $M_{BH} = 0.1\% M_{*}$, where the $\sigma_{o}$ is lower in the central region. This issue is pronounced in isotropic models, which typically exhibit lower $\sigma_{o}$ compared to anisotropic models.

In subsequent sections, we will investigate the internal velocity profiles of these models and examine how regularization and varying initial conditions can potentially enhance BH mass recovery, especially in scenarios lacking $h_{3}$ and $h_{4}$.

\subsection{Models with varying regularization} 

The regularization factor plays a crucial role in smoothing the model and avoiding overfitting; however, excessive smoothing sometimes can lead to biased solutions \citep{valluri_etal_2004}. \cite{Lipka.2021} introduced an optimization method for regularization based on the estimation of the complexity of the Schwarzschild models. This method was tested solely on axisymmetric models, assessing its impact on determining the inclination angle and $M/L$. \cite{Pilawa.2024} extended this optimization to triaxial Schwarzschild models with six free parameters, including BH mass, and found that adding a penalty term to the likelihood measure either had little effect or, in some cases, impaired the recovery of BH mass. However, none of the aforementioned studies considered a realistic N-body representation of mock models with various types of velocity anisotropy. 
Furthermore, these studies relied on full kinematic information and did not explore BH recovery in scenarios lacking $h_{3}$, and $h_{4}$. 

In this section, we explore the effect of different regularization values tested on our isotropic and anisotropic models with and without higher-order velocity moments. 


As presented in Fig. \ref{fig:chinoh34}, for both isotropic and anisotropic models that incorporate higher-order velocity moments $h_3$, and $h_4$ (blue contours), increasing $\lambda$ leads to a narrowing of the likelihood surface. Nevertheless, the one-sigma interval still recovers the true value of the BH mass for all $\lambda$ values.

In contrast, for models lacking $h_{3}$ and $h_{4}$ constraints (red contours), the choice of high or low regularization values becomes crucial depending on the model's isotropy. This is because, without employing $h_3$, and $h_4$, we are essentially unable to accurately recover the true isotropy profile of the target. We found that adopting higher regularization leads to a more isotropic model. Therefore, for isotropic mock models without $h_3$, and $h_4$, increasing  $\lambda$ helps in accurately recovering the black hole mass as it leads to more isotropic models (first row in Fig. \ref{fig:chinoh34}). 

Conversely, in anisotropic models without $h_3$, and $h_4$, lower regularization yields a more accurate black hole mass recovery, whereas high regularization biases the results significantly by overestimating BH mass and underestimating the $M/L$ (second row in Fig. \ref{fig:chinoh34}). While previous research
suggests UCDs generally display isotropic characteristics \citep[]{seth2014}, we found that when we lack $h_{3}$ and $h_{4}$ constraints, setting $\lambda=20-30$ still leads to relatively good recovery for both isotropic and anisotropic mock UCDs, but this holds true for only models where the BH mass is $M_{BH} = 10 \% M_{*}$, which have a significant influence on the $\sigma_{o}$ in the central region. However, models with $M_\mathrm{BH}=$1\%$M_*$ and $M_{BH} = 0.1\% M_{*}$ that lack $h_{3}$ and $h_{4}$ constraints, continue to show significant bias, regardless of $\lambda$ values.

\begin{figure*}
	\centering	%
	\includegraphics[width=1\columnwidth]{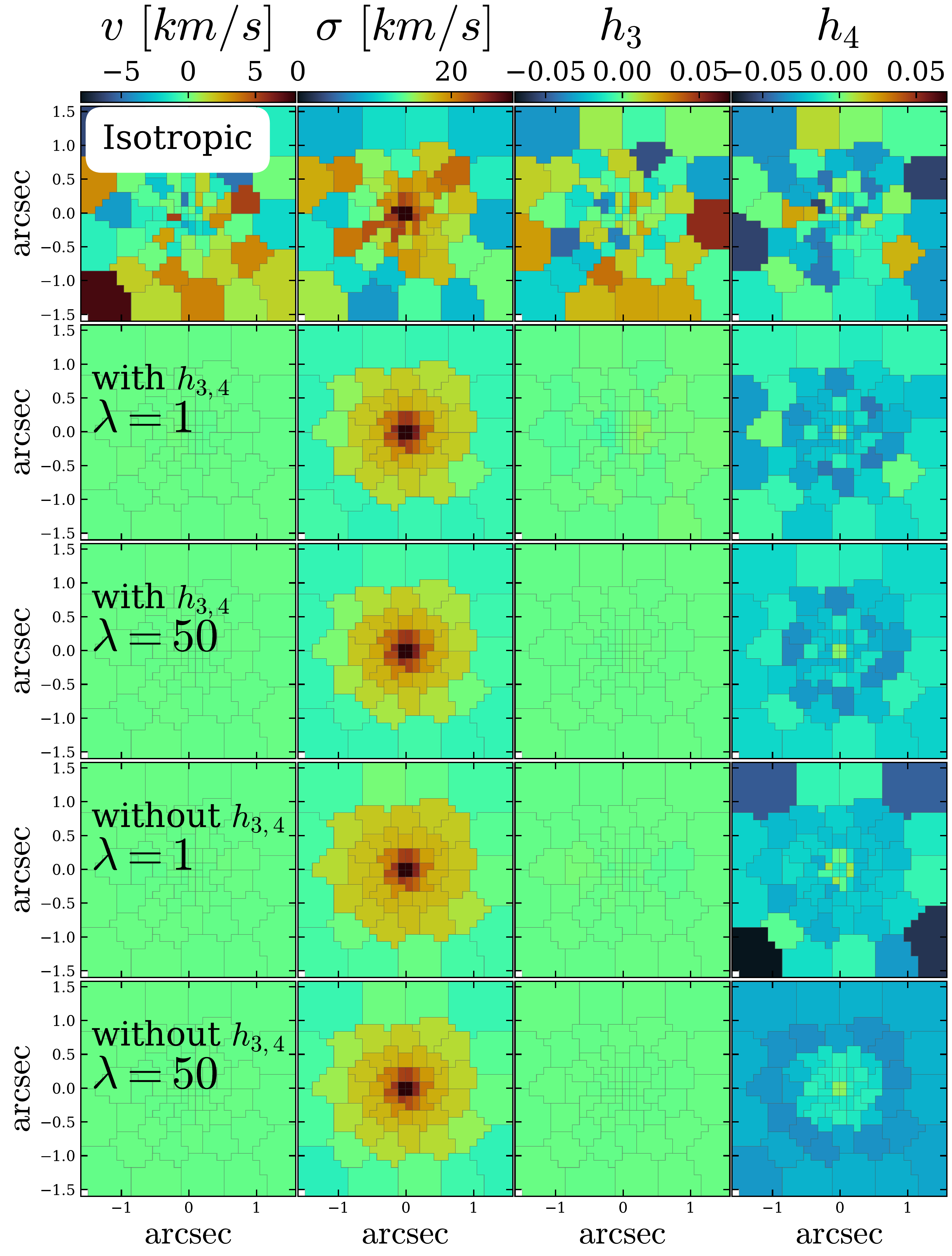}
 	\includegraphics[width=1\columnwidth]{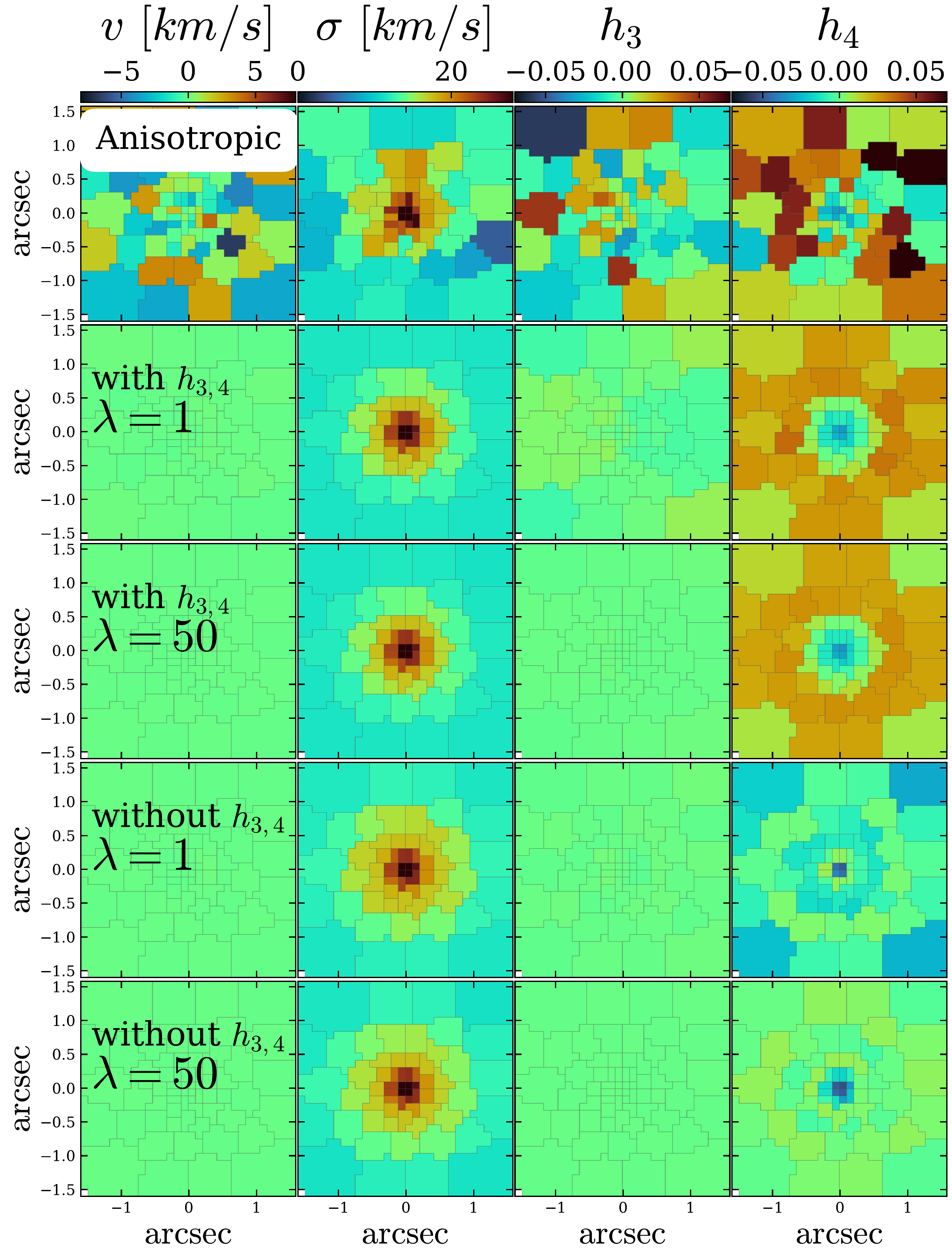}
  \caption{Top row: noise-added kinematic maps for an isotropic (left panels) and anisotropic (right panels) mock UCDs with a BH mass of $M_{\rm BH} = 10\% M_*$.  The second and third rows show the Schwarzschild models that explicitly fit $h_3$, and $h_4$ constraints with regularization values of $\lambda = 1$ and $\lambda = 50$, respectively. The fourth and fifth rows show Schwarzschild models without using $h_3$, and $h_4$ constraints with $\lambda = 1$ and $\lambda = 50$, respectively. }%
	\label{fig:models}%
\end{figure*}

Fig. \ref{fig:models} shows the kinematic map examples of our Schwarzschild models for an isotropic and anisotropic mock UCD with a BH mass of $M_{\rm BH} = 10\% M_*$.  The first row shows the noise-added kinematic maps and the other rows show the kinematic maps resulting from  Schwarzschild models with and without higher-order velocity moments constraints for $\lambda$ values of 1 and 50.  Models are plotted for the true value of BH and $M/L$. Although the $h_{3}$ and $h_{4}$ values are not directly constrained in the models shown in the last two rows, we computed them to compare with models that use $h_{3}$ and $h_{4}$ constrains.

Models with $h_{3}$ and $h_{4}$ constrains fit both the $\sigma$ and $h_{4}$ maps well, and there is not much difference between the models with $\lambda=1$ and $\lambda=50$. In models without $h_{3}$ and $h_{4}$ constraints, the $\sigma$ maps are well fitted for both $\lambda=1$ and $\lambda=50$, however, the $h_{4}$ maps show more negative values across the entire field of view, implying a more tangential velocity distribution.

\begin{figure*}
	\centering	
	\includegraphics[width=1\columnwidth]{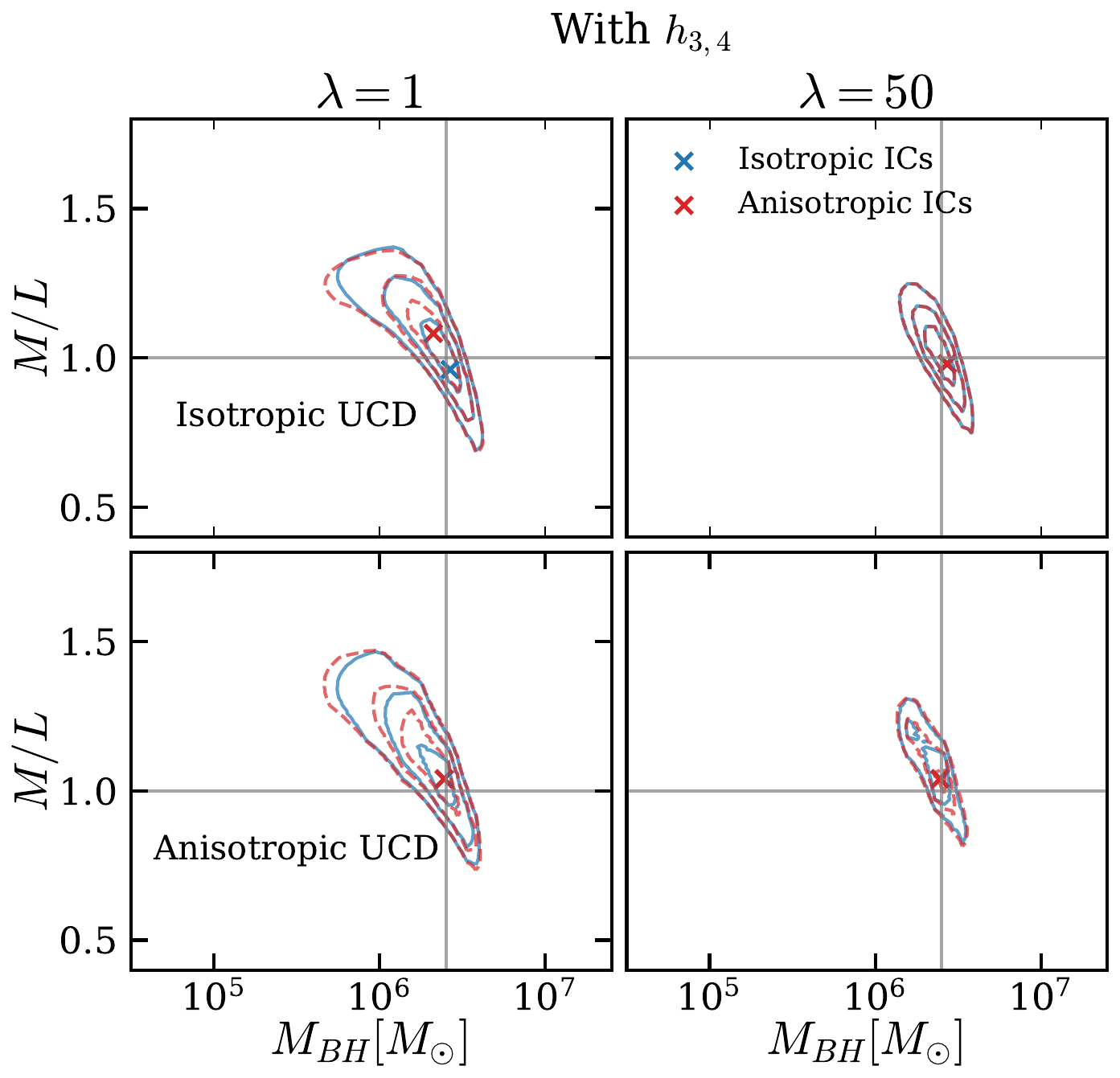}
 	\includegraphics[width=1\columnwidth]{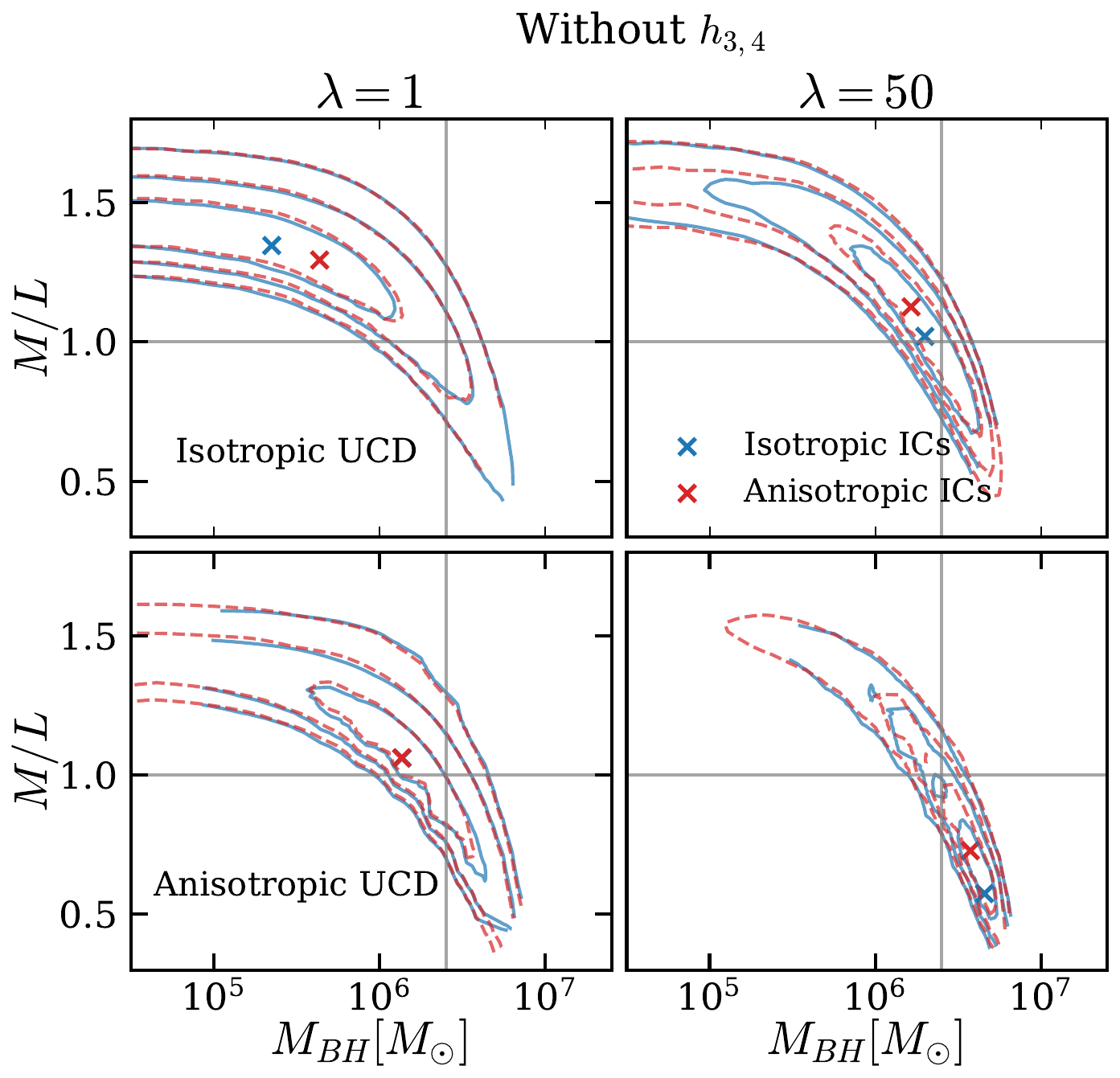}
	\hspace{8pt}%
	\caption{Comparing the results of modeling the
 isotropic UCD (top row) and anisotropic UCD (bottom row) with $M_{BH} = 10 \% M_{*}$, using higher-order GH moments $h_3$, $h_4$ (left panels) and using only $v$, $\sigma$ (right panels), with isotropic initial conditions (blue contours) and anisotropic initial conditions (red contours). The first and second columns of each panel show the results of models with regularization factors of $\lambda=1$, and $50$, respectively.}%
	\label{fig:chinoh34_IC}%
\end{figure*}

\begin{figure*}
	\centering	
	\includegraphics[width=1\columnwidth]{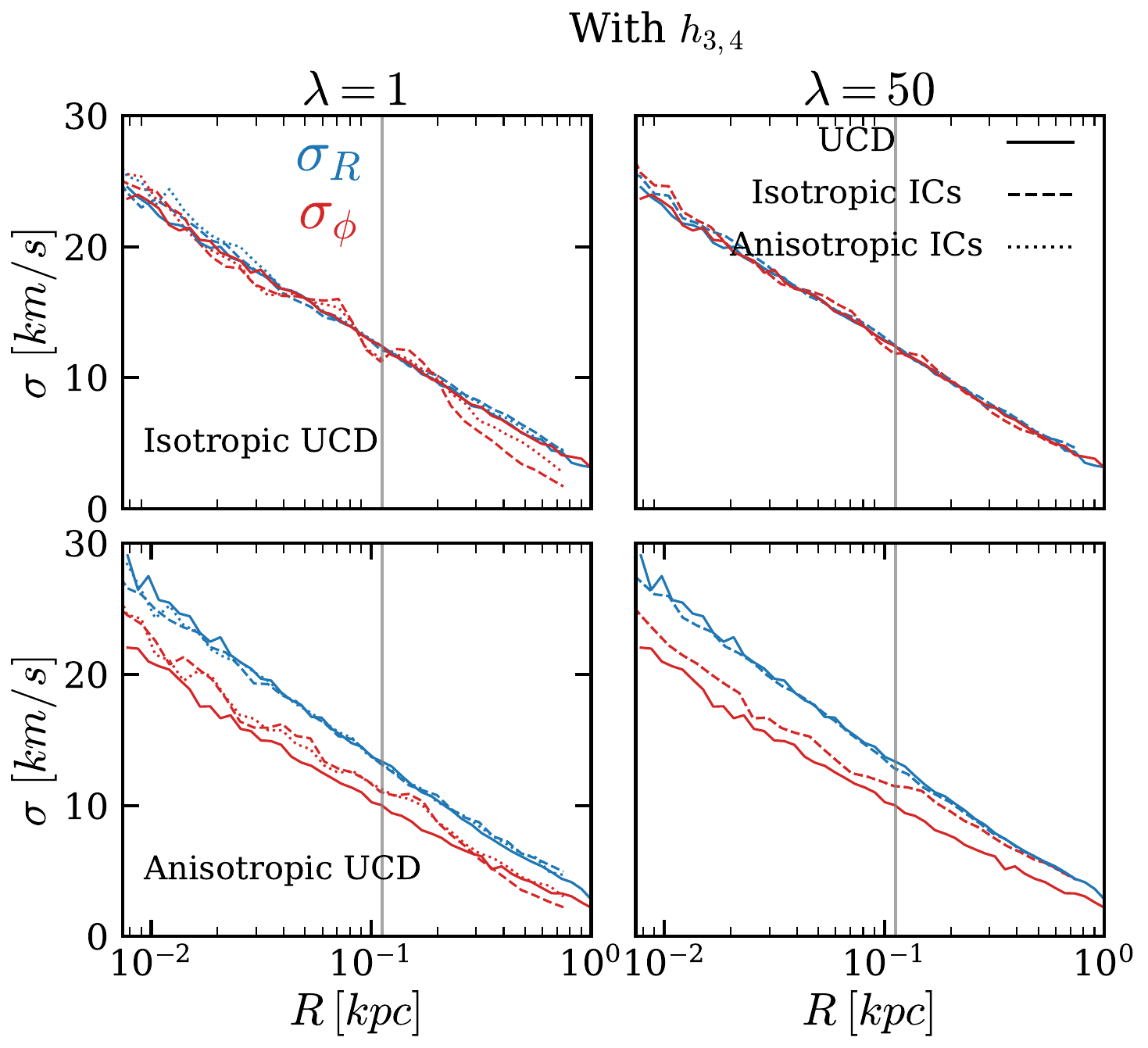}
 	\includegraphics[width=1\columnwidth]{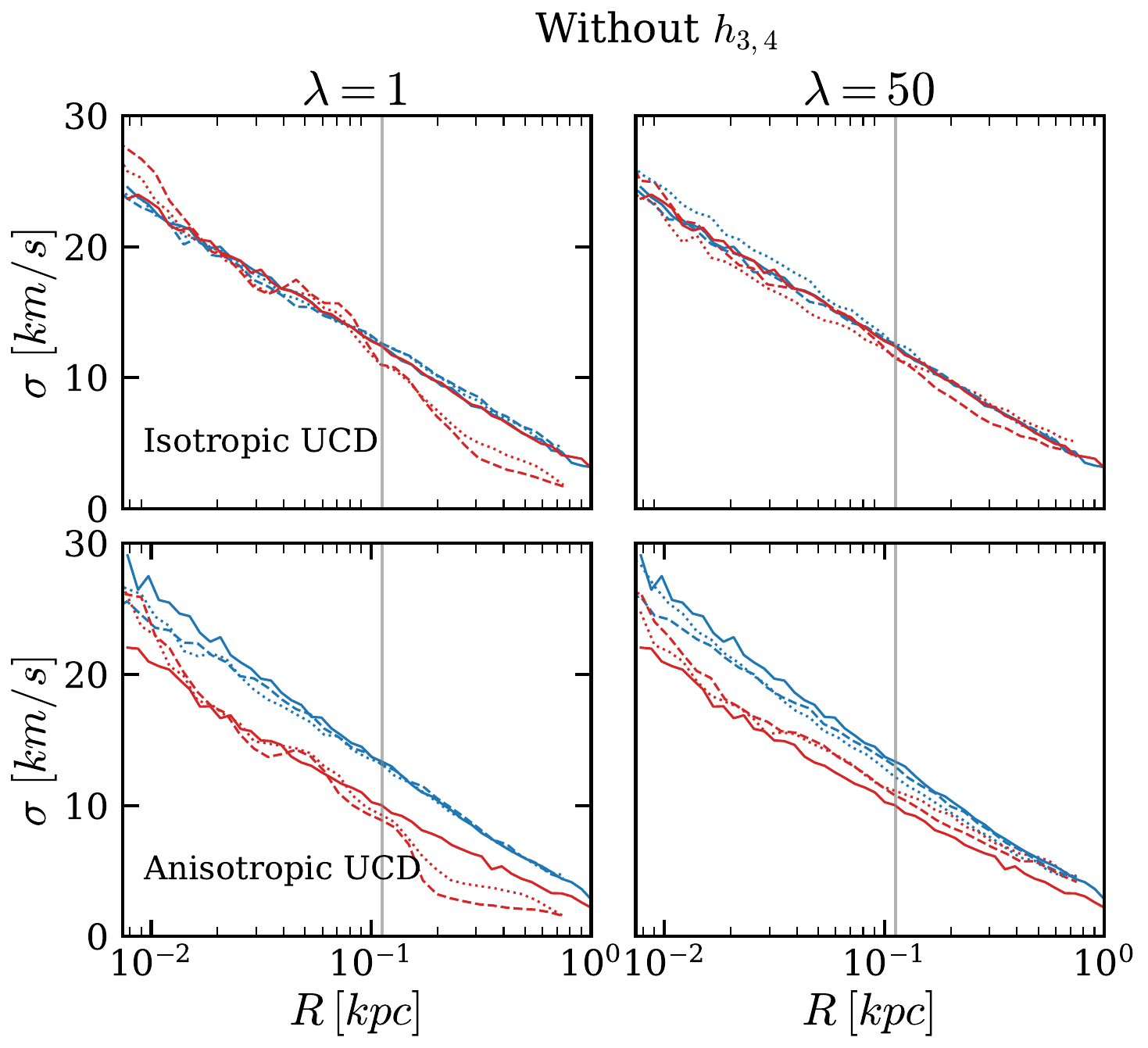}
	\hspace{8pt}%
	\caption{Comparing the intrinsic velocity dispersions $\sigma_{R}$ (in blue) and $\sigma_{\phi}$ (in red) in models presented in Fig. \ref{fig:chinoh34_IC}. Solid lines represent the dispersion profiles of the mock UCDs. Models with isotropic initial conditions are shown by dashed lines, while those with anisotropic initial conditions are indicated by dotted lines. All the models are plotted for the true values of $M_{BH}$ and $M/L$. The vertical gray line indicates the extent of our kinematic mock data.}%
	\label{fig:intrinsic_vel}%
\end{figure*}

\subsection{Modeling with various initial conditions}

As discussed, in cases where the target is anisotropic, modeling without employing $h_{3}$ and $h_{4}$ constraints, and with high regularization, can bias the recovery of BH mass. High regularization tends to produce a more isotropic model. However, all the models we have previously discussed are constructed using orbit libraries derived from isotropic initial conditions, regardless of whether the target UCD is isotropic or anisotropic.

To investigate potential improvements in BH recovery in anisotropic UCDs with a high regularization parameter, we construct orbit libraries with various sets of initial conditions generated by different anisotropy parameters. We facilitate this by setting a non-zero value of anisotropy parameter $\beta_0$ during the assignment of velocities to randomly chosen positions within the given density profile. As described in \ref{IC_detail}, this process involves solving the Jeans equation for the axisymmetrized potential and density. For both isotropic and anisotropic UCDs, we sampled initial conditions in two scenarios: initially, considering $\beta_0 = 0$ to achieve an isotropic solution for the Jeans equation, and subsequently, using $\beta_0 = 0.5$, which yields an anisotropic solution. We then reran our models with regularization factors of $\lambda = 1$ and $50$.


Fig. \ref{fig:chinoh34_IC} shows the modeling results for isotropic UCD in the top row and anisotropic UCD in the bottom row with $M_{BH} = 10\% M_{*}$, employing $h_{3}$ and $h_{4}$ constraints in the left panels and excluding these constraints in the right panels, with different regularization settings of $\lambda =1$ and $50$. Models created with isotropic initial conditions are depicted in blue, while those with anisotropic initial conditions are shown in red. We found that models with both isotropic and anisotropic initial conditions yield similar outcomes, including for models with $M_{BH} = 1\% M_{*}$ and $M_{BH} = 0.1\% M_{*}$. However, in the absence of $h_{3}$ and $h_{4}$ constraints, assigning isotropic or anisotropic initial conditions can result in slight variations in the best-fitting values. Despite adopting various initial conditions, the impact of high regularization continues to be significant, yielding more isotropic model outcomes, even when initial conditions are generated with a high anisotropy parameter $\beta=0.5$.



Fig. \ref{fig:intrinsic_vel} shows the intrinsic radial velocity dispersions $\sigma_{R}$ (in blue) and tangential velocity dispersions $\sigma_{\phi}$ of models presented in Fig. \ref{fig:chinoh34_IC}.  Solid lines represent the velocity dispersion profiles of the mock UCDs. Models with isotropic initial conditions are shown by dashed lines, while those with anisotropic initial conditions are indicated by dotted lines. The vertical gray line indicates the extent of our kinematic mock data. 

In the first row of the right panels, the isotropic model without $h_{3}$ and $h_{4}$ constraints, shows significant deviations in $\sigma_{R}$ and $\sigma_{\phi}$ from an isotropic velocity distribution. However, employing higher regularization closely aligns the model with the mock's intrinsic velocity.

The second row of the right panels demonstrates that high regularization can lead to deviations in the Schwarzschild model velocity distribution from that of the mock anisotropic UCD, especially beyond the extent of the kinematic data. Conversely, lower regularization aligns the model's intrinsic velocity distribution more closely with the anisotropic mock UCD.



\subsection{Sensitivity of the recovered BH mass to $S/N$ and spatial PSF}
 
\begin{figure*}
	\centering	
	\includegraphics[width=2\columnwidth]{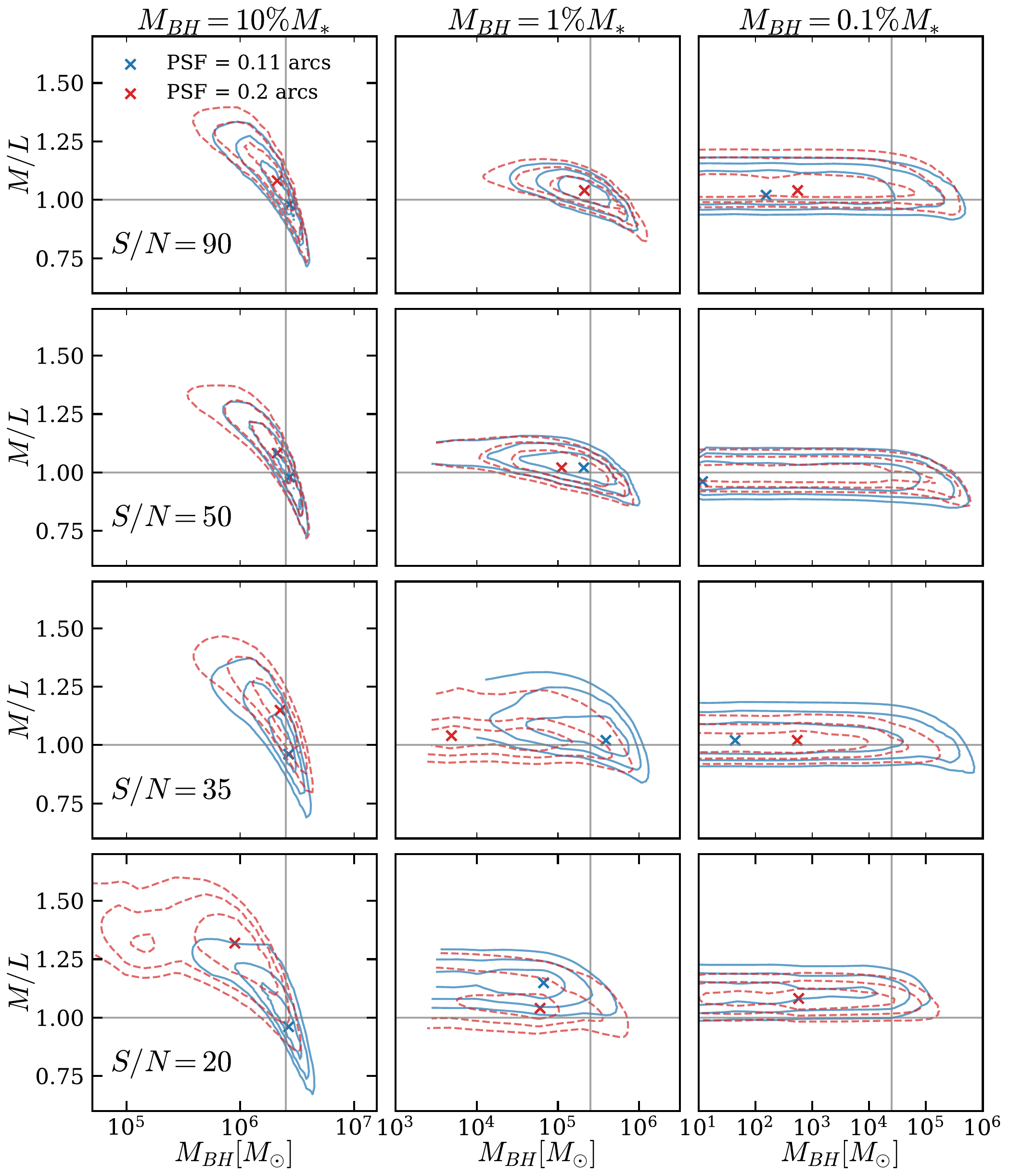}
	\hspace{8pt}%
	\caption{Comparing the results of modeling the isotropic UCD with $M_{BH} = 10 \% M_{*}$, $M_{BH} = 1\% M_{*}$, and $M_{BH} = 0.1\% M_{*}$, by generating kinematic maps with various $S/N$ and spatial PSF sizes. Each row from top to bottom represents results for $S/N$ values of 90, 50, 35, and 20. Contours show the results of modeling with a well-resolved PSF (FWHM = 0.11 arcseconds) in blue, and with a relatively large PSF size (FWHM = 0.2 arcseconds) in red.}%
	\label{fig:SN}%
\end{figure*}

We examine the ability of our modeling approach to recover different BH masses under various  conditions for generating mock kinematic: considering different observational conditions such as $S/N$ and spatial PSF. Fig. \ref{fig:SN} shows the results of modeling the isotropic UCD using kinematic maps generated with $S/N$ of 90, 50, 35, and 20. Blue contours represent models with a well-resolved PSF with FWHM $\sim 0.11$ arcseconds, while red contours indicate models with a larger PSF size with FWHM $\sim 0.2$ arcseconds.

For our mock models, the BH with  $M_{BH} = 0.1\% M_{*}$ can not be recovered, even at high $S/N$ and using higher moment constraints up to $h_8$ (see Appendix \ref{appen2}). This is expected, as the BH sphere of influence for this model is smaller than the spatial resolution, which we consider throughout this study. As shown in Fig. \ref{fig:Sigma}, there is almost no difference in the intrinsic velocity dispersion profile of the model with a BH of $M_{BH} = 0.1\% M_{*}$ and the model without a BH.
However, we emphasize that the inability to recover BH masses $M_{BH} \leq 0.1\% M_{*}$ in the Virgo cluster does not imply such mass fractions cannot be detected by  JWST/NIRSpec IFU in galaxies that are nearer to us. This detection limit is also specific to our spherical, non-rotating mocks described in Section \ref{mock_gen} (although this is a reasonable assumption for UCDs). The fractional BH mass detectable could be different for more complicated models that incorporate rotation with axisymmetric or triaxial shapes, or different stellar profiles, all of which can affect the stellar distribution function and the location of the BH sphere of influence.

We demonstrate that when the PSF is well-resolved at 0.11 arcsec, a BH with a mass of $M_{BH} = 10 \% M_{*}$ can be reliably determined with a $S/N$ of at least 15. However, for a BH with $M_{BH} = 1\% M_{*}$, a higher  $S/N \gtrsim 30$ is needed for accurate recovery. As shown in Figure \ref{fig:SN}, with a larger PSF of 0.2 arcsec and $S/N=20$, all recovered BH masses exhibit significant biases. For this larger PSF, a BH mass of $M_{BH} = 10 \% M_{*}$ can be accurately recovered with $S/N \gtrsim 30$, while a BH mass of $M_{BH} = 1\% M_{*}$ requires $S/N \gtrsim 45$ for accurate recovery.

\begin{figure*}
	\centering	
	\includegraphics[width=2\columnwidth]{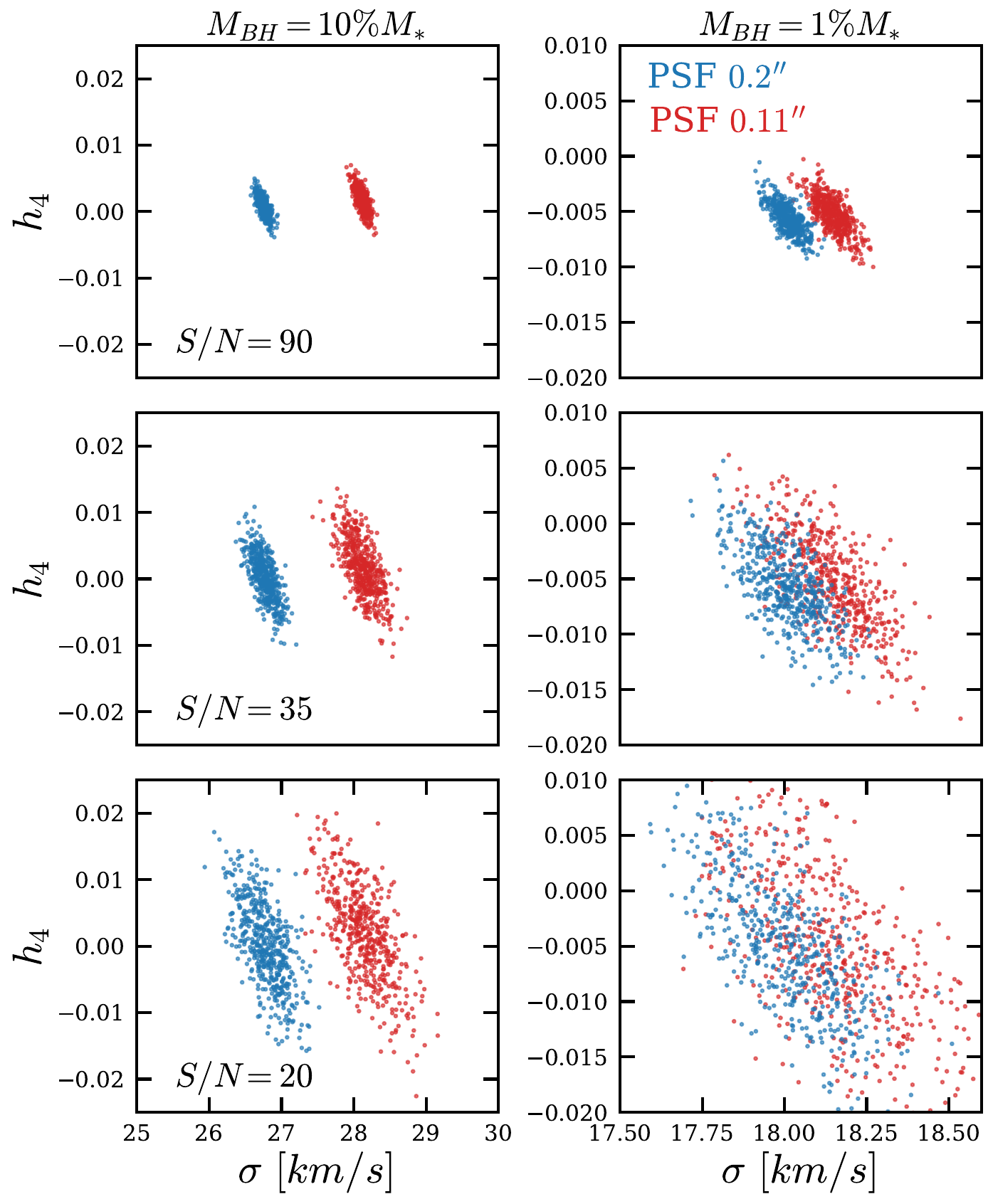}
	\hspace{8pt}%
	\caption{Variations in kinematic moments in the central Voronoi bin extracted from N-body representation of 500 mock isotropic UCDs created with different random seeds, with $M_{BH} = 10 \% M_{*}$ (left column), and $M_{BH} = 1\% M_{*}$ (right column). For each random seed, the spatial grid is convolved once with a PSF of 0.11 arcseconds (red dots) and once with a PSF of 0.2 arcseconds (blue dots). Rows from top to bottom represent the results for different $S/N$ ratios of 90, 35, and 20. The larger PSF leads to an overall smaller $\sigma$, with reductions of up to 6\% for models with $M_{BH} = 10 \% M_{*}$, and 1\% for models with $M_{BH} = 1 \% M_{*}$. }%
	\label{fig:psfkin}%
\end{figure*}

We note that for the larger PSF, $M/L$ of the best-fit models is systematically overestimated by $\sim 10\%$ for all mock UCDs with $M_{BH} = 10 \% M_{*}$ and $S/N \gtrsim 30$ , although the true value still falls within the 1$\sigma$ uncertainty range. This overestimation of $M/L$ can be explained by Fig. \ref{fig:psfkin}, which illustrates the impact of PSF size on the extracted kinematic moments in the central Voronoi bin derived from a set of 500 mock isotropic UCDs generated with varied random seeds, featuring $M_{BH} = 10 \% M_{*}$ (left column) and $M_{BH} = 1\% M_{*}$ (right column). Each mock realization is convolved twice: once with a PSF of 0.11 arcseconds (red dots) and once with a PSF of 0.2 arcseconds (blue dots). The rows, from top to bottom, correspond to different $S/N$ of 90, 35, and 20.
As anticipated, a mild degeneracy exists between $h_{4}$ and $\sigma$ across various N-body realizations, which does not significantly impact our modeling outcomes. However, the larger PSF results in a systematically smaller $\sigma$ up to $\sim 6-8\%$ for models with $M_{BH} = 10 \% M_{*}$ and $\sim 1-2\%$ for models with $M_{BH} = 1 \% M_{*}$, while it does not significantly affect the $h_{4}$ values. The effect of a larger PSF on the $\sigma$ is due to the averaging of velocities over a broader area, resulting in a reduced width of the LOSVD. In contrast, $h_{4}$, which measures the shape (peakedness) of the LOSVD, is a higher-order moment and is less affected by the PSF smoothing and averaging effects. Consequently, to compensate for the decrease of $\sigma$ across the kinematic map, $M/L$ is overestimated by $\sim 10\%$, particularly at lower $S/N$, where the overestimation is more pronounced. The recovery of the BH mass is less affected and is underestimated by less than $5\%$, meaning the signature of the BH on the shape of the LOSVD, particularly in the central regions, remains detectable even after larger PSF convolution, as long as the PSF is smaller than the sphere of influence of the BH.

\section{Conclusions} \label{sec:cite}

In this study, we explored the 
kinematics of UCDs and cEs containing a central BH by constructing mock N-body realizations. Then, we generated the spatially resolved stellar kinematics of these systems at a distance of the Virgo cluster under a similar resolution as JWST/NIRSpec IFU. Utilizing the orbit-superposition code FORSTAND, we constructed dynamical models for our mock sample. This methodology has been efficient in establishing a potential lower limit for the detectable SMBH masses in compact stellar systems in the Virgo cluster using JWST/NIRSpec IFU.

We find that black hole masses constituting less than $1\%$ of the stellar mass do not affect the intrinsic velocity dispersion enough to be detectable in the central region, since the black hole sphere of influence falls below the spatial resolution of JWST/NIRspec IFU. The impact of black hole masses on the velocity dispersion profile becomes notable when the black hole mass exceeds a few percent of the stellar mass. This effect is more pronounced in anisotropic models than in isotropic ones. Additionally, the sphere of influence of the black hole is more extensive in cEs than in UCDs.

We found that, in our case, the deprojection method does not significantly affect the accuracy of black hole mass estimation. This may be due to the NIRSpec/IFU spatial resolution being comparable to or larger than that of HST images. If the IFU's resolution is significantly lower than that of the HST images, deprojecting a Sersic profile to a cuspy profile could potentially lead to an underestimation of the black hole mass. However, in the future dynamical modeling of real systems, it will be beneficial to employ both Sersic profile deprojection and the MGE parametrization to get robust measurements. Overall, we find that a black hole with a mass of a few percent $(> 1 \%)$ of the total host stellar mass is recoverable using full kinematic constraints.

We also highlight the challenges and limitations in dynamical modeling when higher-order moments of the line-of-sight velocity distribution are not considered. This limitation underscores the importance of incorporating these higher-order moments to overcome the mass-anisotropy degeneracy problem. 

Our findings indicate that, without 
$h_{3}$ and $h_{4}$ constraints, higher regularization inherently produces more isotropic models. This approach yields more accurate results for isotropic models, but for anisotropic targets, it can significantly bias the outcomes.

Through extensive testing, we determined that without employing $h_{3}$ and $h_{4}$ constraints, only an intermediate value of regularization parameter allows for the reliable recovery of BH mass in both isotropic and anisotropic models, and it only holds true for models with $M_{BH} = 10\% M_{*}$. Intermediate regularization values, neither too high nor too low, help balance the model's performance, regardless of the target's isotropy characteristics.



\begin{acknowledgments}
We thank Michael Drinkwater for  helpful discussions. BT, AL, and MV acknowledge funding from Space Telescope Science Institute awards: JWST-GO-02567.002-A and HST-GO-16882.002-A. EV acknowledges support from an STFC Ernest Rutherford fellowship (ST/X004066/1). M.A.T.\ and S.T.\ acknowledge the support of the Canadian Space Agency (CSA) [22JWGO1-07]. 
\end{acknowledgments}

\appendix




\section{A Compact Elliptical Model Including Dark Matter} \label{appen}
\begin{figure*}
	\centering	%
\includegraphics[width=\columnwidth]{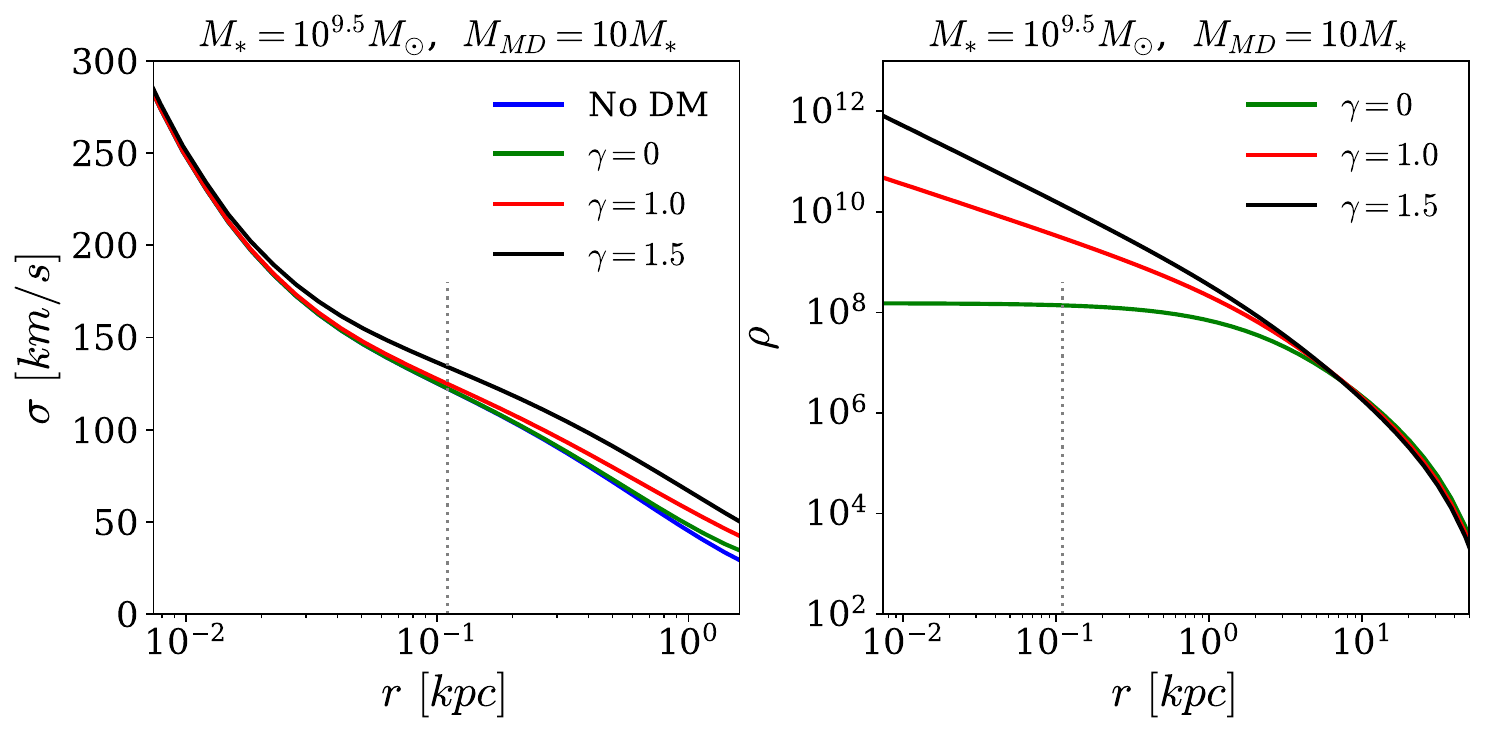}
	\caption{Left: Density profile of spherical DM models with $\gamma = 0$, corresponding to a cored profile (green), $\gamma = 1$ equivalent to the NFW profile (red), and $\gamma = 1.5$ representing a very cuspy model. Right: intrinsic velocity dispersion profiles of mock isotropic cEs with $M_{BH} = 10\% M_{*}$, incorporating each of the DM models from the left panel. The blue line represents a cE without DM. The vertical dashed lines in each panel mark the extent of the mock kinematic data.}%
	\label{fig:DM_S}%
\end{figure*}

\begin{figure*}
	\centering	%
\includegraphics[width=1\columnwidth]{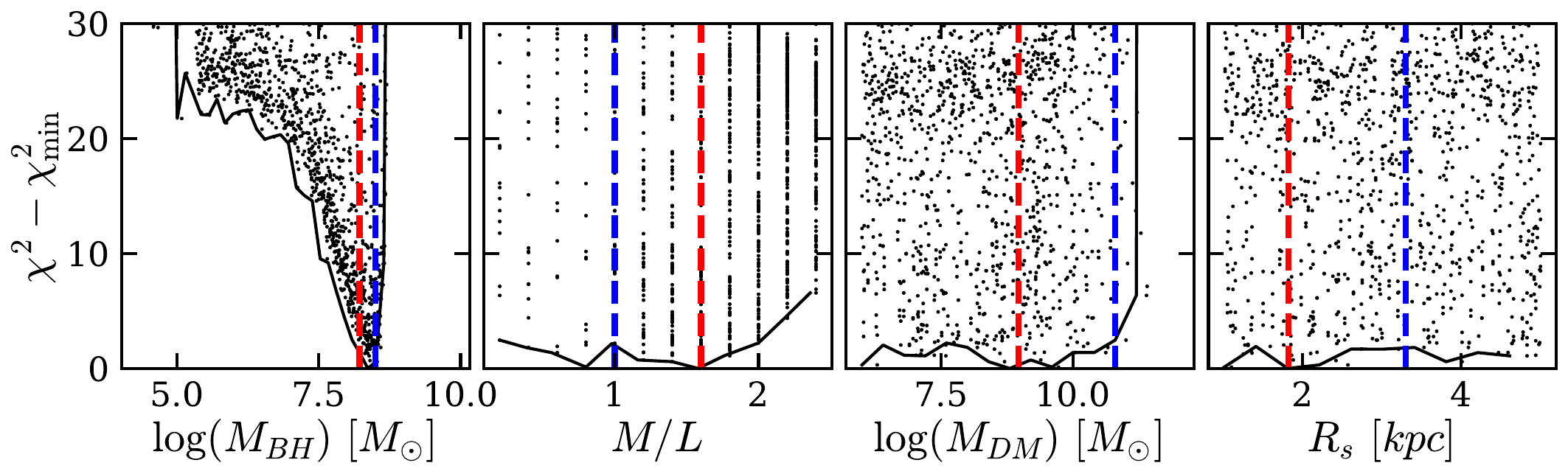}
\includegraphics[width=0.72\columnwidth]{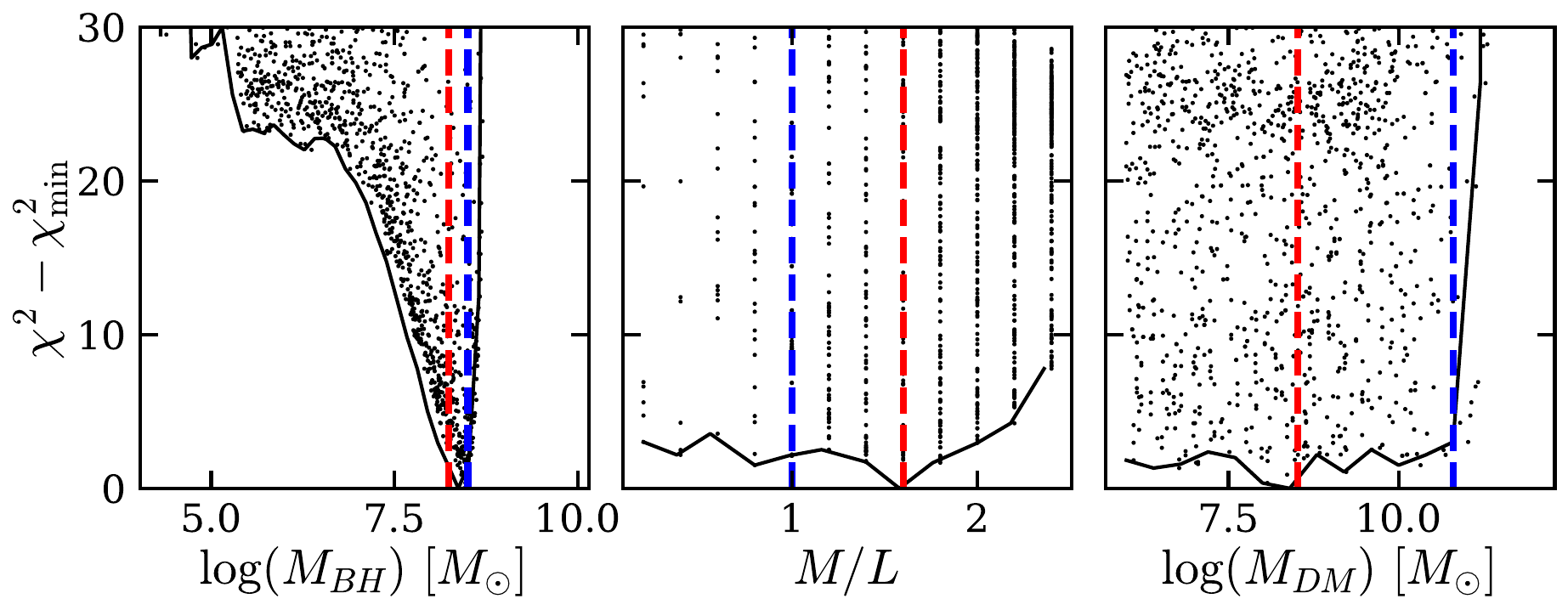}	
\caption{The top row shows the results of modeling a mock isotropic cE with four free parameters, including the DM component. Panels from left to right display $\chi^{2}-\chi^{2}_{min}$ as functions of BH mass, mass-to-light ratio, DM mass, and DM scale radius, respectively. Each point represents a model. The vertical red dashed lines mark the parameters of the best-fitting model, while the blue dashed lines indicate the true values of parameters in the mock cE. The bottom row shows the modeling of the same mock cE when we fix the DM scale radius to the true value of 3.3 kpc.}%
	\label{fig:DM}%
\end{figure*}

We explore a cE model that includes a DM halo. The DM matter is constructed using a similar approach with spherical density (eq. \ref{eq:1}), with parameters as $a = 3.3$ kpc, $M_* = 2\times 10^{10.5}$ $M_{\odot}$, $\beta = 3$, $\alpha = 1$, and $r_{\mathrm{cut}} = 10a$. We adopt various $\gamma$ values to generate DM profiles with different inner slopes. Three DM models are generated: $\gamma = 0$, corresponding to a cored profile; $\gamma = 1$, corresponding to an NFW profile; and $\gamma = 1.5$, giving a density profile cuspier than NFW as shown in the left panel in Fig. \ref{fig:DM_S}. The right panel in Fig. \ref{fig:DM_S} shows the effect of each DM model on the total intrinsic velocity dispersion of the cE compared to a cE model without DM. Although the DM constitutes $90\%$ of the cE, cored and NFW profiles do not change the intrinsic velocity dispersion profile within our kinematic extent (see \citet{Borriello.2003} for additional discussion). Only the DM halo with a very cuspy profile causes some changes in the inner region of the cE. Thus, we construct a dynamical model based on a cE with a very cuspy DM profile. The free parameters include the BH mass, $M/L$, the DM mass ($M_{DM}$), and the DM halo scale radius $R_s$. We conducted model searches using the efficient method of Latin hypercube sampling \cite[]{McKay}. The variation of $\chi^{2}-\chi^{2}_{min}$ with each parameters are presented in Fig. \ref{fig:DM}. The top row shows the results of modeling with four free parameters, while the bottom row shows the results when we fix the DM scale radius to the true value of 3.3 kpc. The DM parameters cannot be constrained due to the limited data coverage of kinematics. Even when we fix the DM scale radius, there is a significant degeneracy between the $M/L$ and the DM mass. Despite this, the BH mass can still be reasonably recovered with $15\%$ accuracy.

\section{Modeling with higher order moment constraints
} \label{appen2}
\begin{figure*}
	\centering	
	\includegraphics[width=\columnwidth]{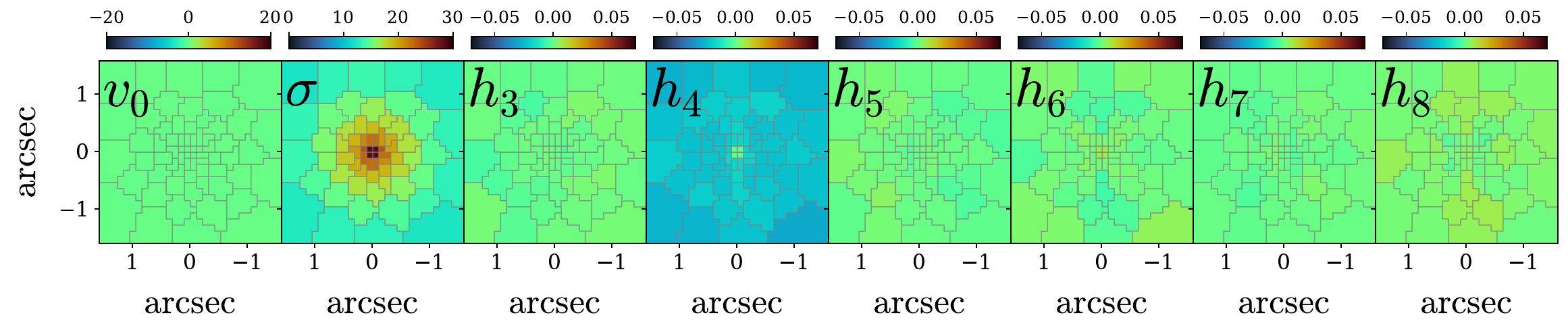}
	\hspace{8pt}%
	\caption{The noise-free kinematic maps for a mock isotropic UCDs with $M_{BH} = 10\% M_{*}$ up to 8th-order GH coefficients.}%
	\label{fig:h6h8map}%
\end{figure*}

\begin{figure*}
	\centering	
	\includegraphics[width=\columnwidth]{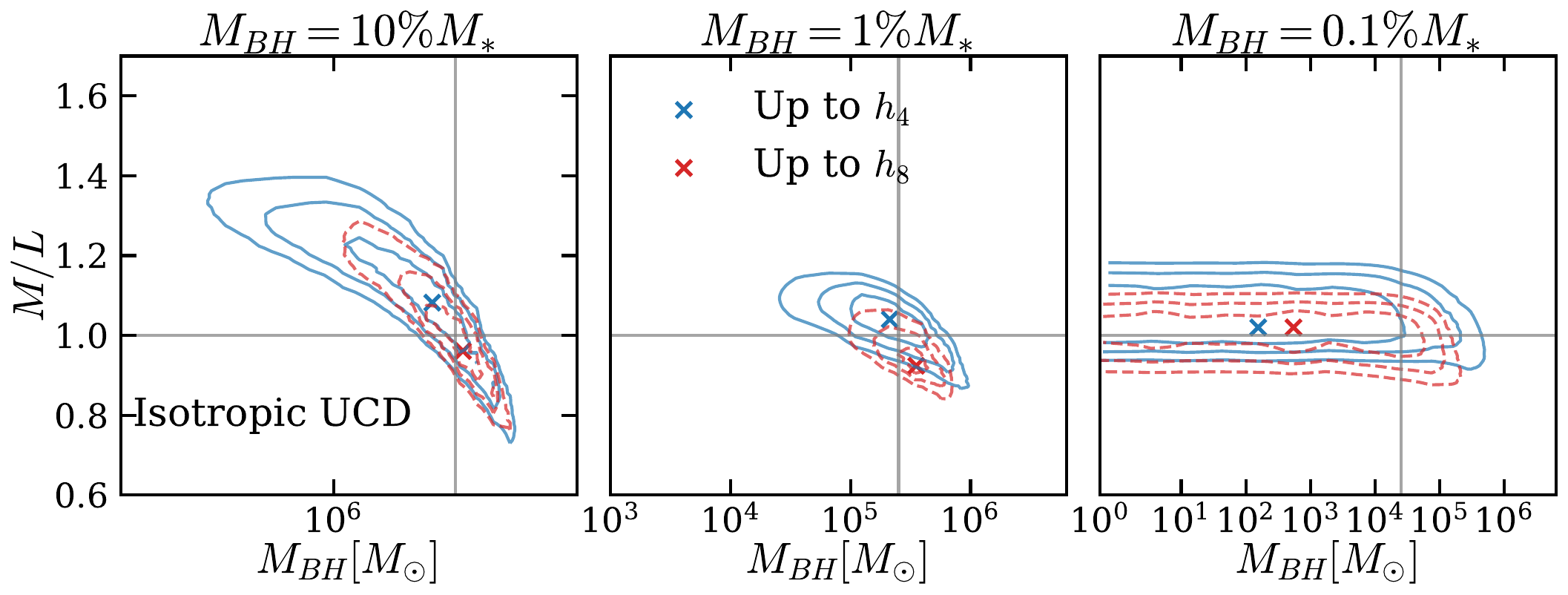}
	\hspace{8pt}%
	\caption{Comparing model results for an isotropic UCD with $M_{BH} = 10 \% M_{*}$, $M_{BH} = 1\% M_{*}$, and $M_{BH} = 0.1\% M_{*}$, using higher order moment constraints up to $h_{4}$ (blue contours) and $h_{8}$ (red contours).}%
	\label{fig:h6h8}%
\end{figure*}

Fig. \ref{fig:h6h8map} represents the noise-free kinematic maps with up to 8th-order GH coefficients for a mock isotropic UCDs with $M_{BH} = 10\% M_{*}$. The kinematic maps are generated with $S/N=90$ as such higher order velocity moments are only reliably recoverable with high $S/N$. The signals in $h_{6}$ and $h_{8}$ are weak which is probably due to the simplicity of our model. Fig. \ref{fig:h6h8} compares the recovery of true $M_{BH}$ and $M/L$, using higher-order moment kinematics up to $h_{4}$ (blue contours) versus up to $h_{8}$ (red contours). We see that utilizing higher order moments leads to the shrinkage of contours and tighter constraints for models with $M_{BH} = 10 \% M_{*}$, and  $M_{BH} = 1\% M_{*}$. This is primarily due to the increased number of kinematic constraints (since both sets of models same number of orbits) and does not necessarily imply that the higher order moments help to improve/tighten the constraints. Once again we see that an  $M_{BH} = 0.1\% M_{*}$ is not recoverable due to the unresolved sphere of influence of the BH, even with higer order moments.

\bibliography{sample631}{}

\begin{thebibliography}{}
\expandafter\ifx\csname natexlab\endcsname\relax\def\natexlab#1{#1}\fi
\providecommand{\url}[1]{\href{#1}{#1}}
\providecommand{\dodoi}[1]{doi:~\href{http://doi.org/#1}{\nolinkurl{#1}}}
\providecommand{\doeprint}[1]{\href{http://ascl.net/#1}{\nolinkurl{http://ascl.net/#1}}}
\providecommand{\doarXiv}[1]{\href{https://arxiv.org/abs/#1}{\nolinkurl{https://arxiv.org/abs/#1}}}

\bibitem[{{Afanasiev} {et~al.}(2018){Afanasiev}, {Chilingarian}, {Mieske}, {Voggel}, {Picotti}, {Hilker}, {Seth}, {Neumayer}, {Frank}, {Romanowsky}, {Hau}, {Baumgardt}, {Ahn}, {Strader}, {den Brok}, {McDermid}, {Spitler}, {Brodie}, \& {Walsh}}]{Afanasiev.2018}
{Afanasiev}, A.~V., {Chilingarian}, I.~V., {Mieske}, S., {et~al.} 2018, \mnras, 477, 4856, \dodoi{10.1093/mnras/sty913}

\bibitem[{{Ahn} {et~al.}(2017){Ahn}, {Seth}, {den Brok}, {Strader}, {Baumgardt}, {van den Bosch}, {Chilingarian}, {Frank}, {Hilker}, {McDermid}, {Mieske}, {Romanowsky}, {Spitler}, {Brodie}, {Neumayer}, \& {Walsh}}]{ahn2017}
{Ahn}, C.~P., {Seth}, A.~C., {den Brok}, M., {et~al.} 2017, \apj, 839, 72, \dodoi{10.3847/1538-4357/aa6972}

\bibitem[{{Ahn} {et~al.}(2018){Ahn}, {Seth}, {Cappellari}, {Krajnovi{\'c}}, {Strader}, {Voggel}, {Walsh}, {Bahramian}, {Baumgardt}, {Brodie}, {Chilingarian}, {Chomiuk}, {den Brok}, {Frank}, {Hilker}, {McDermid}, {Mieske}, {Neumayer}, {Nguyen}, {Pechetti}, {Romanowsky}, \& {Spitler}}]{Ahn.2018}
{Ahn}, C.~P., {Seth}, A.~C., {Cappellari}, M., {et~al.} 2018, \apj, 858, 102, \dodoi{10.3847/1538-4357/aabc57}

\bibitem[{{Bekki} {et~al.}(2001){Bekki}, {Couch}, {Drinkwater}, \& {Gregg}}]{Bekki.2001}
{Bekki}, K., {Couch}, W.~J., {Drinkwater}, M.~J., \& {Gregg}, M.~D. 2001, \apjl, 557, L39, \dodoi{10.1086/323075}

\bibitem[{{Bekki} {et~al.}(2003){Bekki}, {Couch}, {Drinkwater}, \& {Shioya}}]{2003MNRAS.344..399B}
{Bekki}, K., {Couch}, W.~J., {Drinkwater}, M.~J., \& {Shioya}, Y. 2003, \mnras, 344, 399, \dodoi{10.1046/j.1365-8711.2003.06916.x}

\bibitem[{{Binney} \& {Mamon}(1982)}]{Binney.1982}
{Binney}, J., \& {Mamon}, G.~A. 1982, \mnras, 200, 361, \dodoi{10.1093/mnras/200.2.361}

\bibitem[{{Borriello} {et~al.}(2003){Borriello}, {Salucci}, \& {Danese}}]{Borriello.2003}
{Borriello}, A., {Salucci}, P., \& {Danese}, L. 2003, \mnras, 341, 1109, \dodoi{10.1046/j.1365-8711.2003.06404.x}

\bibitem[{{Brodie} {et~al.}(2011){Brodie}, {Romanowsky}, {Strader}, \& {Forbes}}]{Brodie_2011}
{Brodie}, J.~P., {Romanowsky}, A.~J., {Strader}, J., \& {Forbes}, D.~A. 2011, \aj, 142, 199, \dodoi{10.1088/0004-6256/142/6/199}

\bibitem[{{Buote} \& {Barth}(2019)}]{Buote.2019}
{Buote}, D.~A., \& {Barth}, A.~J. 2019, \apj, 877, 91, \dodoi{10.3847/1538-4357/ab1008}

\bibitem[{{Cappellari}(2002)}]{cappellari.2002}
{Cappellari}, M. 2002, \mnras, 333, 400, \dodoi{10.1046/j.1365-8711.2002.05412.x}

\bibitem[{{Cappellari}(2008)}]{Cappellari.2008}
---. 2008, \mnras, 390, 71, \dodoi{10.1111/j.1365-2966.2008.13754.x}

\bibitem[{{Cappellari} \& {Copin}(2003)}]{Cappellari.2003}
{Cappellari}, M., \& {Copin}, Y. 2003, \mnras, 342, 345, \dodoi{10.1046/j.1365-8711.2003.06541.x}

\bibitem[{{Chilingarian} {et~al.}(2009){Chilingarian}, {Cayatte}, {Revaz}, {Dodonov}, {Durand}, {Durret}, {Micol}, \& {Slezak}}]{Chilingarian.2009S}
{Chilingarian}, I., {Cayatte}, V., {Revaz}, Y., {et~al.} 2009, Science, 326, 1379, \dodoi{10.1126/science.1175930}

\bibitem[{{Crain} {et~al.}(2015){Crain}, {Schaye}, {Bower}, {Furlong}, {Schaller}, {Theuns}, {Dalla Vecchia}, {Frenk}, {McCarthy}, {Helly}, {Jenkins}, {Rosas-Guevara}, {White}, \& {Trayford}}]{crain_eagle}
{Crain}, R.~A., {Schaye}, J., {Bower}, R.~G., {et~al.} 2015, \mnras, 450, 1937, \dodoi{10.1093/mnras/stv725}

\bibitem[{{Cuddeford}(1991)}]{Cuddeford.1991}
{Cuddeford}, P. 1991, \mnras, 253, 414, \dodoi{10.1093/mnras/253.3.414}

\bibitem[{{Deeley} {et~al.}(2023){Deeley}, {Drinkwater}, {Sweet}, {Bekki}, {Couch}, \& {Forbes}}]{Deeley.2023}
{Deeley}, S., {Drinkwater}, M.~J., {Sweet}, S.~M., {et~al.} 2023, \mnras, 525, 1192, \dodoi{10.1093/mnras/stad2313}

\bibitem[{{Drinkwater} {et~al.}(2000){Drinkwater}, {Jones}, {Gregg}, \& {Phillipps}}]{Drinkwater.2000}
{Drinkwater}, M.~J., {Jones}, J.~B., {Gregg}, M.~D., \& {Phillipps}, S. 2000, \pasa, 17, 227, \dodoi{10.1071/AS00034}

\bibitem[{{Fellhauer} \& {Kroupa}(2002)}]{Fellhauer.2002}
{Fellhauer}, M., \& {Kroupa}, P. 2002, \mnras, 330, 642, \dodoi{10.1046/j.1365-8711.2002.05087.x}

\bibitem[{{Frank} {et~al.}(2011){Frank}, {Hilker}, {Mieske}, {Baumgardt}, {Grebel}, \& {Infante}}]{Frank.2011}
{Frank}, M.~J., {Hilker}, M., {Mieske}, S., {et~al.} 2011, \mnras, 414, L70, \dodoi{10.1111/j.1745-3933.2011.01058.x}

\bibitem[{{Gerhard}(1993)}]{Gerhard.1993}
{Gerhard}, O.~E. 1993, \mnras, 265, 213, \dodoi{10.1093/mnras/265.1.213}

\bibitem[{{Gregg} {et~al.}(2003){Gregg}, {Drinkwater}, {Hilker}, {Phillipps}, {Jones}, \& {Ferguson}}]{2003Ap&SS.285..113G}
{Gregg}, M.~D., {Drinkwater}, M.~J., {Hilker}, M.~J., {et~al.} 2003, \apss, 285, 113, \dodoi{10.1023/A:1024622128955}

\bibitem[{{Hilker} {et~al.}(1999){Hilker}, {Infante}, {Vieira}, {Kissler-Patig}, \& {Richtler}}]{Hilker.1999}
{Hilker}, M., {Infante}, L., {Vieira}, G., {Kissler-Patig}, M., \& {Richtler}, T. 1999, Astron. Astrophys. Suppl. Ser., 134, 75, \dodoi{10.1051/aas:1999434}

\bibitem[{{Lipka} \& {Thomas}(2021)}]{Lipka.2021}
{Lipka}, M., \& {Thomas}, J. 2021, \mnras, 504, 4599, \dodoi{10.1093/mnras/stab1092}

\bibitem[{{Liu} {et~al.}(2020){Liu}, {C{\^o}t{\'e}}, {Peng}, {Roediger}, {Zhang}, {Ferrarese}, {S{\'a}nchez-Janssen}, {Guhathakurta}, {Yang}, {Jing}, {Alamo-Mart{\'\i}nez}, {Blakeslee}, {Boselli}, {Cuilandre}, {Duc}, {Durrell}, {Gwyn}, {Jord{\'a}n}, {Ko}, {Lan{\c{c}}on}, {Lim}, {Longobardi}, {Mei}, {Mihos}, {Mu{\~n}oz}, {Powalka}, {Puzia}, {Spengler}, \& {Toloba}}]{liu_ngvs}
{Liu}, C., {C{\^o}t{\'e}}, P., {Peng}, E.~W., {et~al.} 2020, \apjs, 250, 17, \dodoi{10.3847/1538-4365/abad91}

\bibitem[{{Mayes} {et~al.}(2024){Mayes}, {Drinkwater}, {Pfeffer}, \& {Baumgardt}}]{mayes_2024}
{Mayes}, R., {Drinkwater}, M., {Pfeffer}, J., \& {Baumgardt}, H. 2024, \mnras, 527, 4643, \dodoi{10.1093/mnras/stad3428}

\bibitem[{{Mayes} {et~al.}(2021){Mayes}, {Drinkwater}, {Pfeffer}, {Baumgardt}, {Liu}, {Ferrarese}, {C{\^o}t{\'e}}, \& {Peng}}]{2021MNRAS.506.2459M}
{Mayes}, R.~J., {Drinkwater}, M.~J., {Pfeffer}, J., {et~al.} 2021, \mnras, 506, 2459, \dodoi{10.1093/mnras/stab1864}

\bibitem[{McKay {et~al.}(1979)McKay, Beckman, \& Conover}]{McKay}
McKay, M.~D., Beckman, R.~J., \& Conover, W.~J. 1979, Technometrics, 21, 239.
\newblock \url{http://www.jstor.org/stable/1268522}

\bibitem[{{Merrell} {et~al.}(2023){Merrell}, {Vasiliev}, {Bentz}, {Valluri}, \& {Onken}}]{Merrell.2023}
{Merrell}, K.~A., {Vasiliev}, E., {Bentz}, M.~C., {Valluri}, M., \& {Onken}, C.~A. 2023, \apj, 949, 13, \dodoi{10.3847/1538-4357/acc4bc}

\bibitem[{{Merrifield} \& {Kent}(1990)}]{Merrifield&Kent.1990}
{Merrifield}, M.~R., \& {Kent}, S.~M. 1990, \aj, 99, 1548, \dodoi{10.1086/115438}

\bibitem[{{Merritt}(1987)}]{Merritt.1987}
{Merritt}, D. 1987, \apj, 313, 121, \dodoi{10.1086/164953}

\bibitem[{{Mieske} {et~al.}(2013){Mieske}, {Frank}, {Baumgardt}, {L{\"u}tzgendorf}, {Neumayer}, \& {Hilker}}]{mieske2013}
{Mieske}, S., {Frank}, M.~J., {Baumgardt}, H., {et~al.} 2013, \aap, 558, A14, \dodoi{10.1051/0004-6361/201322167}

\bibitem[{{Mieske} {et~al.}(2006){Mieske}, {Hilker}, {Infante}, \& {Jord{\'a}n}}]{Mieske.2006}
{Mieske}, S., {Hilker}, M., {Infante}, L., \& {Jord{\'a}n}, A. 2006, \aj, 131, 2442, \dodoi{10.1086/500583}

\bibitem[{{Mieske} {et~al.}(2012){Mieske}, {Hilker}, \& {Misgeld}}]{mieske_2012}
{Mieske}, S., {Hilker}, M., \& {Misgeld}, I. 2012, A\&A, 537, A3, \dodoi{10.1051/0004-6361/201117634}

\bibitem[{{Navarro} {et~al.}(1996){Navarro}, {Frenk}, \& {White}}]{NFW.1996}
{Navarro}, J.~F., {Frenk}, C.~S., \& {White}, S. D.~M. 1996, \apj, 462, 563, \dodoi{10.1086/177173}

\bibitem[{{Nguyen} {et~al.}(2019){Nguyen}, {Seth}, {Neumayer}, {Iguchi}, {Cappellari}, {Strader}, {Chomiuk}, {Tremou}, {Pacucci}, {Nakanishi}, {Bahramian}, {Nguyen}, {den Brok}, {Ahn}, {Voggel}, {Kacharov}, {Tsukui}, {Ly}, {Dumont}, \& {Pechetti}}]{Nguyen.2019}
{Nguyen}, D.~D., {Seth}, A.~C., {Neumayer}, N., {et~al.} 2019, \apj, 872, 104, \dodoi{10.3847/1538-4357/aafe7a}

\bibitem[{{Peng} {et~al.}(2010){Peng}, {Ho}, {Impey}, \& {Rix}}]{Peng.2010}
{Peng}, C.~Y., {Ho}, L.~C., {Impey}, C.~D., \& {Rix}, H.-W. 2010, \aj, 139, 2097, \dodoi{10.1088/0004-6256/139/6/2097}

\bibitem[{{Pilawa} {et~al.}(2024){Pilawa}, {Liepold}, \& {Ma}}]{Pilawa.2024}
{Pilawa}, J., {Liepold}, E.~R., \& {Ma}, C.-P. 2024, arXiv e-prints, arXiv:2403.07996, \dodoi{10.48550/arXiv.2403.07996}

\bibitem[{{Roberts} {et~al.}(2021){Roberts}, {Bentz}, {Vasiliev}, {Valluri}, \& {Onken}}]{Roberts.2021}
{Roberts}, C.~A., {Bentz}, M.~C., {Vasiliev}, E., {Valluri}, M., \& {Onken}, C.~A. 2021, \apj, 916, 25, \dodoi{10.3847/1538-4357/ac05b6}

\bibitem[{{Schaye} {et~al.}(2015){Schaye}, {Crain}, {Bower}, {Furlong}, {Schaller}, {Theuns}, {Dalla Vecchia}, {Frenk}, {McCarthy}, {Helly}, {Jenkins}, {Rosas-Guevara}, {White}, {Baes}, {Booth}, {Camps}, {Navarro}, {Qu}, {Rahmati}, {Sawala}, {Thomas}, \& {Trayford}}]{schaye_eagle}
{Schaye}, J., {Crain}, R.~A., {Bower}, R.~G., {et~al.} 2015, \mnras, 446, 521, \dodoi{10.1093/mnras/stu2058}

\bibitem[{{Schwarzschild}(1979)}]{sch_1979}
{Schwarzschild}, M. 1979, \apj, 232, 236, \dodoi{10.1086/157282}

\bibitem[{{Seth} {et~al.}(2010){Seth}, {Cappellari}, {Neumayer}, {Caldwell}, {Bastian}, {Olsen}, {Blum}, {Debattista}, {McDermid}, {Puzia}, \& {Stephens}}]{Seth.2010}
{Seth}, A.~C., {Cappellari}, M., {Neumayer}, N., {et~al.} 2010, \apj, 714, 713, \dodoi{10.1088/0004-637X/714/1/713}

\bibitem[{{Seth} {et~al.}(2014){Seth}, {van den Bosch}, {Mieske}, {Baumgardt}, {Brok}, {Strader}, {Neumayer}, {Chilingarian}, {Hilker}, {McDermid}, {Spitler}, {Brodie}, {Frank}, \& {Walsh}}]{seth2014}
{Seth}, A.~C., {van den Bosch}, R., {Mieske}, S., {et~al.} 2014, \nat, 513, 398, \dodoi{10.1038/nature13762}

\bibitem[{{Smith} {et~al.}(2016){Smith}, {Choi}, {Lee}, {Rhee}, {Sanchez-Janssen}, \& {Yi}}]{Smith.2016}
{Smith}, R., {Choi}, H., {Lee}, J., {et~al.} 2016, \apj, 833, 109, \dodoi{10.3847/1538-4357/833/1/109}

\bibitem[{{Strader} {et~al.}(2013){Strader}, {Seth}, {Forbes}, {Fabbiano}, {Romanowsky}, {Brodie}, {Conroy}, {Caldwell}, {Pota}, {Usher}, \& {Arnold}}]{Strader.2013}
{Strader}, J., {Seth}, A.~C., {Forbes}, D.~A., {et~al.} 2013, \apjl, 775, L6, \dodoi{10.1088/2041-8205/775/1/L6}

\bibitem[{{Thomas} {et~al.}(2008){Thomas}, {Drinkwater}, \& {Evstigneeva}}]{2008MNRAS.389..102T}
{Thomas}, P.~A., {Drinkwater}, M.~J., \& {Evstigneeva}, E. 2008, \mnras, 389, 102, \dodoi{10.1111/j.1365-2966.2008.13543.x}

\bibitem[{{Tsatsi} {et~al.}(2015){Tsatsi}, {Macci{\`o}}, {van de Ven}, \& {Moster}}]{Tsatsi.2015}
{Tsatsi}, A., {Macci{\`o}}, A.~V., {van de Ven}, G., \& {Moster}, B.~P. 2015, \apjl, 802, L3, \dodoi{10.1088/2041-8205/802/1/L3}

\bibitem[{{Valluri} {et~al.}(2004){Valluri}, {Merritt}, \& {Emsellem}}]{valluri_etal_2004}
{Valluri}, M., {Merritt}, D., \& {Emsellem}, E. 2004, \apj, 602, 66, \dodoi{10.1086/380896}

\bibitem[{{van der Marel} \& {Franx}(1993)}]{Marel.1993}
{van der Marel}, R.~P., \& {Franx}, M. 1993, \apj, 407, 525, \dodoi{10.1086/172534}

\bibitem[{{Vasiliev}(2019)}]{agama}
{Vasiliev}, E. 2019, \mnras, 482, 1525, \dodoi{10.1093/mnras/sty2672}

\bibitem[{{Vasiliev} \& {Valluri}(2020)}]{Vasiliev.2020A}
{Vasiliev}, E., \& {Valluri}, M. 2020, \apj, 889, 39, \dodoi{10.3847/1538-4357/ab5fe0}

\bibitem[{{Verolme} {et~al.}(2002){Verolme}, {Cappellari}, {Copin}, {van der Marel}, {Bacon}, {Bureau}, {Davies}, {Miller}, \& {de Zeeuw}}]{Verolme.2002}
{Verolme}, E.~K., {Cappellari}, M., {Copin}, Y., {et~al.} 2002, \mnras, 335, 517, \dodoi{10.1046/j.1365-8711.2002.05664.x}

\bibitem[{{Voggel} {et~al.}(2019){Voggel}, {Seth}, {Baumgardt}, {Mieske}, {Pfeffer}, \& {Rasskazov}}]{voggel2019}
{Voggel}, K.~T., {Seth}, A.~C., {Baumgardt}, H., {et~al.} 2019, \apj, 871, 159, \dodoi{10.3847/1538-4357/aaf735}

\bibitem[{{Y{\i}ld{\i}r{\i}m} {et~al.}(2017){Y{\i}ld{\i}r{\i}m}, {van den Bosch}, {van de Ven}, {Mart{\'\i}n-Navarro}, {Walsh}, {Husemann}, {G{\"u}ltekin}, \& {Gebhardt}}]{Yildirim.2017}
{Y{\i}ld{\i}r{\i}m}, A., {van den Bosch}, R. C.~E., {van de Ven}, G., {et~al.} 2017, \mnras, 468, 4216, \dodoi{10.1093/mnras/stx732}

\bibitem[{{Zhang} {et~al.}(2015){Zhang}, {Peng}, {C{\^o}t{\'e}}, {Liu}, {Ferrarese}, {Cuillandre}, {Caldwell}, {Gwyn}, {Jord{\'a}n}, {Lan{\c{c}}on}, {Li}, {Mu{\~n}oz}, {Puzia}, {Bekki}, {Blakeslee}, {Boselli}, {Drinkwater}, {Duc}, {Durrell}, {Emsellem}, {Firth}, \& {S{\'a}nchez-Janssen}}]{Zhang.2015}
{Zhang}, H.-X., {Peng}, E.~W., {C{\^o}t{\'e}}, P., {et~al.} 2015, \apj, 802, 30, \dodoi{10.1088/0004-637X/802/1/30}

\bibitem[{{Zolotov} {et~al.}(2015){Zolotov}, {Dekel}, {Mandelker}, {Tweed}, {Inoue}, {DeGraf}, {Ceverino}, {Primack}, {Barro}, \& {Faber}}]{Zolotov.2015}
{Zolotov}, A., {Dekel}, A., {Mandelker}, N., {et~al.} 2015, \mnras, 450, 2327, \dodoi{10.1093/mnras/stv740}

\end{thebibliography}
\bibliographystyle{aasjournal}

\end{document}